\documentstyle[aps,pra]{revtex}

\newcommand{\bra}[1]{\langle #1 |\,}
\newcommand{\bbra}[1]{\langle \langle #1 |\,}
\newcommand{\Bra}[1]{\Big\langle #1 \Big|\,}
\newcommand{\ket}[1]{\,| #1 \rangle}
\newcommand{\kket}[1]{\,| #1 \rangle \rangle}
\newcommand{\bracket}[2]{\langle #1 | #2 \rangle}
\newcommand{\bbracket}[2]{\langle \langle #1 | #2 \rangle \rangle}
\newcommand{\brakket}[2]{\langle #1 | #2 \rangle \rangle}
\newcommand{\mat}[1]{\underline{\underline{#1}}}

\newcommand{\bpsi}{\bra{\Psi_0^N}}
\newcommand{\kpsi}{\ket{\Psi_0^N}}

\newcommand{\beq}{\begin{equation}}
\newcommand{\eeq}{\end{equation}}
\newcommand{\beqs}{\begin{eqnarray}}
\newcommand{\eeqs}{\end{eqnarray}}

\newcommand{\ky}[1]{| Y_{#1} \rangle}
\newcommand{\by}[1]{\langle Y_{#1} |}
\newcommand{\By}[1]{\Big\langle Y_{#1} \Big |}

\newcommand{\R}{{\bbox{R}}}
\newcommand{\r}{{\bbox{r}}}
\newcommand{\He}{{H_{\text{e}}}}
\newcommand{\Hez}{{H_{\text{e},0}}}
\newcommand{\Hei}{{H_{\text{e},1}}}
\newcommand{\Hn}{{H_{\text{n}}}}
\newcommand{\bHn}{{\bbox{H}_{\text{n}}}}
\newcommand{\Vo}{{V_0}}
\newcommand{\Tn}{{T_{\text{n}}}}

\newcommand{\Gd}{{\cal{G}}}
\newcommand{\Sd}{{\cal{A}}}  
\newcommand{\Md}{{\cal{M}}}  
\newcommand{\Si}{{\cal{A}}}
\newcommand{\Ti}{{\cal{T}}}
\newcommand{\Gi}{{\cal{G}}}
\newcommand{\Hb}{{\breve{H}}}
\newcommand{\bHb}{{\breve{\bbox{H}}}}
\newcommand{\kpt}{\ket{\Psi_{\text{tot}}(\R)}}
\newcommand{\kptR}{\kpt}
\newcommand{\aop}{{\text{aop}}}

\title{The dynamical Green's function and an exact optical potential
for electron-molecule scattering including nuclear dynamics}

\author{Joachim Brand, Lorenz S. Cederbaum, and Hans-Dieter Meyer}

\address{Theoretische Chemie, Universit\"at Heidelberg, \\
	Im Neuenheimer Feld 229, D-69120 Heidelberg, Germany}

\date{April 9, 1999, {\em acc. for publ.} Phys. Rev. A, physics/9907050}

\begin{document}

\draft

\maketitle


\begin{abstract}
We derive a rigorous optical potential for electron-molecule
scattering including the effects of nuclear dynamics by extending the
common many-body Green's function approach to optical potentials
beyond the fixed-nuclei limit for molecular targets.  Our formalism
treats the projectile electron and the nuclear motion of the target
molecule on the same footing whereby the dynamical optical potential
rigorously accounts for the complex many-body nature of the scattering
target. One central result of the present work is that the common
fixed-nuclei optical potential is a valid adiabatic approximation to
the dynamical optical potential even when projectile and nuclear
motion are (nonadiabatically) coupled as long as the scattering
energy is well below the electronic excitation thresholds of the
target. 
For extremely low projectile velocities, however, when the cross
sections are most sensitive to the scattering potential, we expect the
influences of the nuclear dynamics on the optical potential to become
relevant. For these cases, a
systematic way to improve the adiabatic approximation to
the dynamical optical potential is presented that yields non-local
operators with respect to the nuclear coordinates.
\end{abstract}


\section{Introduction} \label{sec:intro}

The concept of an optical potential as a physical entity that governs
the scattering of a single particle by a composite target is an
intuitively appealing phenomenological concept that goes back to the
early days of nuclear physics. In principle, the scattering of a
nonrelativistic, quantum-mechanical particle off an $N$-particle
target is a many-body problem and governed by the $(N+1)$-particle
Schr\"odinger equation. In the so-called optical model
\cite{feshbach92}, the elastic scattering problem is alternatively
described by an effective single-particle Schr\"odinger (or
Lippmann-Schwinger) equation. All effects of the interaction of the
projectile particle with the target are contained in the so-called
optical potential. In general, this optical potential has to be a very
complicated object: It becomes a {\bf nonlocal} operator because exchange
and rearrangement of target particles have to be considered.
An {\bf energy dependence} has to account for possible excitations of
the target and if inelastic scattering is energetically possible, the
optical potential is {\bf nonhermitian} in order to describe the loss
of scattering amplitude into the inelastic channels. One major
technical advantage of using optical potentials in numerical
calculations is that the scattering problem can be separated from the
many-body problem and the latter can be treated using bound-state
techniques.

The ability of many-body theory to derive exact, though not unique,
optical potentials was a great success in the 1950's
\cite{feshbach58,feshbach62,bell59}. The optical potential of
Feshbach \cite{feshbach92,feshbach58,feshbach62} follows from the most
straightforward derivation by projection from the $(N+1)$-particle
Schr\"odinger equation. An important alternative can be found
in the self energy of the single-particle Green's function of
traditional many-body theory \cite{bell59,csanak71,ring80}.

In contrast to the Green's function approach, the Feshbach optical
potential is not easily amenable to perturbation theoretical
approximations. Achieving a balanced and consistent treatment of
ground- and excited-state correlation is thus difficult to
achieve. Moreover, Mahaux and Sartor \cite{Mahaux91} have found in
applications to nuclear matter that the nonlocality in the Feshbach
optical potential is more complicated than in the Green's function
optical potential. The numerical applications of Feshbach's theory on
electron-molecule scattering reported so far, have been realised by
projection of configuration interaction (CI) matrices
\cite{schneider82,schneider83} but have rarely gone beyond the uncorrelated
Hartree-Fock level for the description of the target's ground
state. The inclusion of correlated, i.~e., multi-configurational
target wavefunctions in this context presents a delicate problem
\cite{lengsfield91,gil93}. 

On the other hand, the Green's
function optical potential \cite{bell59} defined by the Dyson equation
\cite{fetter71} has turned out to be well-behaved and well-suited for
numerical calculations.  The Green's function approach to the optical
potential has the advantage, that the self energy of the
single-particle Green's function is easily accessible through the
Feynman-Dyson perturbation expansion \cite{fetter71}. Thus a balanced
treatment of ground-state and excited-state correlation is
automatically achieved. Various approximation schemes for the self
energies of molecules have been developed
\cite{schneider70,schirmer83,ortiz88} and applications to
electron-molecule scattering have been reported in
Refs.~\cite{klonover77,klonover78,klonover79,berman83,meyer89}.

All the previous calculations employing optical potentials have been
done in the fixed-nuclei limit where the effects of nuclear motion are
neglected altogether.  A great variety of methods and numerical
procedures can be used for solving the single-particle scattering
equations in the fixed-nuclei limit like, e.~g., the R-matrix
\cite{wigner47,gillan87-liste,pfingst94}, the Schwinger-variational
\cite{lucchese86}, the complex-Kohn \cite{miller87,mccurdy87} and
the close-coupling \cite{lane80,morrison95} approaches. An often-used
approximate treatment of the nuclear motion derives from the
Born-Oppenheimer approximation. In this so-called adiabatic-nuclei
approximation
\cite{case56,temkin67,lane80}, fixed-nuclei $S$-matrix elements
for each configuration $\R$ of the atomic nuclei are averaged,
weighted by the nuclear wavefunctions of the isolated target molecule.
This approximation, however, breaks down for low projectile energies
and, in particular, for resonant vibrational excitation (see
e.~g.~\cite{morrison84}).  Finally, in situations where the electronic
projectiles move with velocities that are comparable to the typical
velocities of nuclear motion, it becomes indispensable to treat the
coupled motion correctly.

The most straightforward way of accounting for the full, nonadiabatic
coupling of the nuclear motion to the motion of the projectile
electron is realised in the close-coupling formalism
\cite{lane80,morrison95}.  In this formalism, the scattering
wavefunction of the full problem (i.~e.~the degrees of freedom of
$N+1$ electrons and the nuclear degrees of freedom appear as
variables) is expanded in a finite set of basis functions that are
chosen according to appropriate exact or approximate constants of
motion. Usually it is not feasible to include excited electronic
states of the target in the close-coupling expansion. These states,
however, are necessary to describe polarisation effects correctly. In
the literature, so far model potentials void of rigorous
justification have been used to account for this defect
\cite{lane80,o'connell83,morrison87,morrison93}. The close-coupling
formalism further has the problem that the numerical calculations are
difficult to converge for very low projectile energies
\cite{morrison97}.

An alternative to the one-step close-coupling expansion for treating
the nuclear dynamics in electron-molecule scattering is a two-step
procedure employing the so-called projection-operator (PO) formalism
\cite{cederbaum81,domcke91,cizek98}. Based on the assumption of a
resonance in fixed-nuclei electron scattering, the PO approach can be
used to treat the nuclear dynamics going beyond the adiabatic-nuclei
picture. In the PO formalism the electronic scattering problem is
separated from the nuclear dynamics by Feshbach projection such that
only some resonance parameters, which can be taken from fixed-nuclei
scattering calculations, determine the equations for the nuclear
dynamics that have to be solved. The coupling to the continuum of
electronic scattering wavefunctions enters the nuclear dynamics via a
complex, nonlocal potential. Different levels of approximation for the
nonadiabatic coupling of electronic and nuclear motion are reached by
local and nonlocal models for the complex potential
\cite{cederbaum81}. The PO method has been very successful in
explaining many features of resonant electron-molecule scattering and
it gives a very intuitive picture for the mechanism of many physical
processes like, e.~g., resonant vibrational excitation
\cite{domcke91}. On the other hand it is {\it per se} not an exact
approach because only resonant contributions to the cross sections can
be treated. There are always non-resonant contributions (also called
background contributions) that cannot be evaluated in the PO formalism
but may actually dominate the spectra in off-resonance
scattering. Sometimes no dominating resonance is present at all, and
sometimes numerous resonances contribute which are difficult to
identify individually. For these cases the separation of nuclear and
electronic motion ultimately fails. Also the nonlocality of the
complex potential usually becomes very complicated and difficult to
model for very slow scattering electrons: Details of the
electron-molecule interaction and the dynamical coupling of
electronic, vibrational, and rotational degrees of freedom become
important when the time scales of projectile and nuclear motion in the
target are comparable. As an example we may mention the scattering by
a polar molecule with a supercritical or near-to-critical dipole
moment. As it is well-known, a supercritical, nonrotating dipole ($D >
1.625$ Debye) can bind an excess electron but this bound state is
weakened or even disappears if the dipole is allowed to rotate
\cite{crawford71}. It is, however, very difficult to include the
correct coupling of rotational degrees of freedom in the nonlocal
complex potential model \cite{note-dipole} and thus a full dynamical
calculation including all relevant degrees of freedom may become
necessary in order to describe the physics of very-low energy
scattering by polar molecules correctly.

To our knowledge, no one has so
far given a rigorous derivation of an optical potential for the
scattering electron that takes full account of the nuclear motion. To
close this gap is the main objective of the present communication.  In
this paper we will present an optical potential that is based on the
usual Green's function optical potential but allows to treat
projectile and nuclear motion on the same footing. This so-called
dynamical optical potential is given by the self energy of the
dynamical Green's function, an extension of the usual single-particle
Green's function including the nuclear kinetic-energy operator. The
dynamical optical potential is an optical potential for
electron-molecule scattering that is elastic with respect to
electronic excitations of the target but inelastic with respect to the
vibrations or rotations of the target. The only assumption that is
used throughout this paper is that the Born-Oppenheimer approximation
is valid for the target's ground state. Otherwise the dynamical
optical potential is exact. It takes full account of the many-body
effects of electron scattering including exchange and polarisation as
well as the full nonadiabatic coupling of the projectile motion to the
nuclear motion of the target.  The expressions we derive for the exact
optical potential may provide guidance for the choice of model
potentials on the one hand, and on the other hand are amenable to {\it
ab-initio} approximations and can be calculated with standard
quantum-chemical methods.

This paper is organised as follows: In the next chapter we will
start with reviewing the definition of the traditional single-particle
Green's function of many-body theory. Applied only to the electronic
degrees of freedom of a molecule at fixed nuclear configuration we
also call this propagator the electronic Green's function. In the
subsequent Sec.~\ref{ssec:def-dyn} we then define the dynamical
Green's function, which allows for the treatment of the full molecular
dynamics including the nuclear degrees of freedom as dynamical
variables. We also define the closely related (nuclear) inelastic
Green's function that shows the close relation to the recently
developed theory of electronically inelastic Green's functions.
Sec.~\ref{sec:propagator} is closed by showing how the electronic
propagator can be obtained in a suitable limit from the dynamical
Green's function.  In the third chapter an algebraic derivation of
Dyson's equation for the electronic Green's function is reviewed
before the dynamic Green's function is shown to fulfil a Dyson
equation and the dynamical self energy is defined. Chapter
\ref{sec:scat} treats the relation to electron-molecule
scattering. The dynamical self energy is shown to be closely related
to the optical potential for the coupled motion of the projectile
electron and the atomic nuclei of the target molecule in
electronically elastic scattering. Expressions for the scattering $S$
and $T$ matrices are shown for inelastic processes with respect to the
nuclear coordinates including vibrational or rotational excitation,
associative detachment, and dissociative attachment. In
Sec.~\ref{sec:eff-schr-eq} an effective Schr\"odinger equation using
the dynamical optical potential is discussed. A direct
derivation for the effective Schr\"odinger equation based on
projection operators is given in Sec.~\ref{sec:opt-pot-proj}, which
yields additional insight into the relation of our Green's function
approach to Feshbach's theory. 
The well-known
static-exchange and the fixed-nuclei scattering equations are
identified as approximations to our exact equations in
Secs.~\ref{sec:static-appr} and \ref{sec:fixed-nuclei}.
The rest of Sec.~\ref{sec:approximate-Tn} is devoted to the discussion
of different possible 
approximations to the dynamical optical potential, in particular
concerning the treatment of the nuclear dynamics in virtual electronic
excitations of the target. We also discuss how to compute the dynamical
optical potential using standard {\it ab initio} quantum chemistry methods.


\section{The dynamical single-particle Green's function}
\label{sec:propagator}

Traditional many-body theory \cite{fetter71} deals with systems of
indistinguishable particles. Applications to molecules usually treat
only the electronic degrees of freedom in the many-body formalism and
freeze the nuclear configuration. After briefly reviewing the
definition of this purely electronic single-particle Green's function
we will define the dynamical Green's function, which allows us to
treat the nuclear degrees of freedom as dynamical variables.

\subsection{The electronic Green's function}

The usual single-particle Green's function $G(\r t, \r' t')$ of
non-relativistic many-body theory is defined by \cite{fetter71}
\beqs \label{def-G}
  \nonumber
  i G(\r t, \r' t') &=& \bpsi \overline{\psi}(\r, t)
     \overline{\psi}^\dagger(\r', t') \kpsi \, \theta(t-t') \\
  & & - \bpsi \overline{\psi}^\dagger(\r', t') \overline{\psi}(\r, t)
   \kpsi \, \theta(t'-t) .
\eeqs
Here, $\theta(t)$ is Heavyside's step function, $\kpsi$ denotes the
exact (correlated) ground state of the $N$-particle system, and
$\overline{\psi}^\dagger(\r, t)$ is the Heisenberg field operator,
which creates a particle at position $\r$ and time $t$. Although we
consider fermionic particles, electrons in particular, we suppress the
notation of spin indices for simplicity ($\r$ may be considered as
spin-space variable). In applications of many-body
theory to molecular physics one usually assumes an adiabatic
decoupling of electronic and nuclear motion and thus only treats the
electronic degrees of freedom explicitly. $G(\r t, \r' t')$ is then
the Green's function in a system of electrons moving in the static
potential of the nuclei, fixed at the coordinates $\R$. In the
following, we will call $G(\r t, \r' t')$ the {\bf (purely) electronic
Green's function} in contrast to the later defined dynamical Green's
function, which will allow to treat non-adiabatic coupling of
electronic and nuclear motion.

The Green's function decomposes into two parts, the
so-called particle propagator $G^+$ and hole propagator $G^-$:
\beq
  G(\r t, \r' t') = G^+(\r t, \r' t') + G^-(\r t,
  \r' t') .
\eeq
Noting that the (purely electronic) Heisenberg operator
$\overline{\psi}^\dagger(\r, t)$ evolves in time with the
electronic Hamiltonian $\He$
\beq
  \overline{\psi}^\dagger(\r, t) = e^{i \He t}
  {\psi}^\dagger(\r) e^{-i \He t} ,
\eeq
we can rewrite the particle and the hole propagator in the following
way:
\beqs \label{def-G+}
  iG^+(\r t, \r' t') &=& \bpsi {\psi}(\r) e^{-i[\He
  -\Vo(\R)] (t-t')} {\psi}^\dagger(\r') \kpsi\,\theta(t-t') ,\\
  \label{def-G-}
  iG^-(\r t, \r' t') &=& - \bpsi {\psi}^\dagger(\r') e^{-i[\Vo(\R)
  -\He] (t-t')}{\psi}(\r) \kpsi\,\theta(t'-t) .
\eeqs
Here, we made use of the fact that the electronic ground state $\kpsi$
is an eigenstate of the electronic Hamiltonian $\He$ with the energy
of the electronic ground state $\Vo(\R)$:
\beq \label{def-Vo}
  \He \kpsi = \Vo(\R) \kpsi .
\eeq
We explicitly indicate here the dependence of $\Vo$ on the
nuclear coordinates $\R$ for later use. Of course, $\He$ and $\kpsi$
depend parametrically on $\R$ as well. 

For convenience, we will later discuss the Green's functions mainly in
the frequency instead of the time domain and using an orbital instead
of a position space representation. The corresponding expression for
the electronic Green's function $G_{pq}(\omega)$, which is equivalent
to the definition (\ref{def-G}), reads
\beqs \label{G-pq-omega}
  \nonumber
  G_{pq}(\omega) &=& \bpsi a_p \frac{1}{\omega - \He +
  \Vo(\R) + i\eta} a^\dagger_q \kpsi \\
  && + \bpsi a^\dagger_q
  \frac{1}{\omega - \Vo(\R) + \He - i\eta} a_p \kpsi .
\eeqs
Here, the Fourier transformation into the frequency domain is
defined by ${F}(\omega) =
\int_{-\infty}^\infty d(t\nolinebreak -\nolinebreak t') \,e^{i\omega
(t-t')} F(t - t')$. 
The positive infinitesimal $\eta$ remains from a convergence factor
that has to be introduced to assure the convergence of the Fourier
transformation. The sign of the term $i\eta$ determines the time
ordering by theta functions in Eqs.~(\ref{def-G+}) and (\ref{def-G-})
\cite{fetter71}.
The transformation into the orbital representation is defined by the
transformation of the creation and destruction operators of second
quantisation. For convenience we assume that $\{\varphi_p(\r)\}_p$ is
a discrete but complete set of square-integrable and normalised basis
functions for the single-particle Hilbert space. The creation
operator for an electron in the orbital $\varphi_p(\r)$ is defined by
\beq \label{transf}
  a^\dagger_p = \int d\r\,\varphi_p(\r) \psi^\dagger(\r) .
\eeq
Like usual, the orthonormality of the orbitals implies the canonical
anti-commutation relations $\{a^\dagger_p\,,\,a_q\} = \delta_{pq}$,
etc.

The orbital set has been chosen discrete just for conceptual clearness
and simple notation. The generalisation to continuous sets does not
pose principal problems. Convenient realisations of orbitals may be
found, e.~g., in the Hartree-Fock orbitals of the target (together
with a convenient discretisation of the continuum) or in the momentum
space representation (discretised by placing the system in a finite
box and employing periodic boundary conditions).

\subsection{The dynamical Green's function} \label{ssec:def-dyn}

If we now want to couple the electronic motion dynamically to the
nuclear motion we have to replace the electronic Hamiltonian $\He$ by
the full molecular Hamiltonian 
\beq \label{split-ham}
  H = \Tn + \He .
\eeq
It is given by the sum of the nuclear kinetic energy operator $\Tn$
and the electronic Hamiltonian $\He$.
By convention, $\He$ contains the internuclear repulsion in
addition to the electronic kinetic energy as well as the
electron-nucleus Coulomb attraction and the electron-electron repulsion.
The nuclear kinetic energy operator $\Tn$ generally includes all
vibrational, rotational and translational degrees of freedom for the
target molecule. For convenience we assume that the translational
motion has been separated off. In this case rotation, vibration, and
dissociation are the remaining degrees of freedom for the nuclear
motion.  In any case, $\Tn$ is a differential operator that acts on
the nuclear coordinate $\R$. Thus $\Tn$ does not commute with $\He$ or
$\Vo(\R)$. However, we want to assume in the following 
that the Born-Oppenheimer 
approximation is valid for the electronic ground state of the
molecular target, which implies that the
commutator of $\Tn$ with the electronic ground state $[\Tn\,,\,\kpsi]$
is negligible. Physically, this means that the electronic
configuration in $\kpsi$ depends only weakly on the nuclear
coordinates.

Let us introduce the nuclear Hamiltonian
\beq \label{def-Hn}
  \Hn = \Tn + \Vo(\R)
\eeq
and let $\phi_k(\R)$ be an eigenfunction with eigenvalue $E_{0k}$:
\beq \label{rovib-eigenstates}
  \Hn \phi_k(\R) = E_{0k} \phi_k(\R) .
\eeq
In the following we will also use the abstract Dirac notation
$\ket{\phi_k}$ from time to time for the nuclear eigenstates. The
wavefunction $\phi_k(\R)$ then emerges as the coordinate
representation of the state $\ket{\phi_k}$ by $\phi_k(\R) =
\bracket{\R}{\phi_k}$.  

Under the assumption that $\Tn$ commutes with
$\kpsi$, it is easily seen that the product wavefunction $\kpsi
\phi_k(\R)$ becomes a molecular eigenstate with energy $E_{0k}$:
\beq \label{mol-eigenstates}
  H \kpsi \phi_k(\R) = E_{0k} \kpsi \phi_k(\R) .
\eeq
This Born-Oppenheimer picture is usually an adequate description for
closed-shell molecules in the electronic ground state. The
negligibility of the commutator $[\Tn\,,\,\kpsi]$ is the only
fundamental assumption of the theory we will develop in the
following. At the end of Sec.~\ref{sec:opt-pot-proj} we will return to
discuss the validity of this approximation and show how the theory can
be generalised. At this point we want to stress that we only assume
the Born-Oppenheimer picture to be valid for the nuclear dynamics in
the electronic ground state of the molecule. We neither assume
adiabaticity for the scattering electron nor for the electronically
excited target states!

We now define the {\bf dynamical Green's function} $\Gd(\r t, \r' t')$ as
the sum of the two  components for particle and hole propagation:
\beq \label{Gd-sum}
  \Gd(\r t, \r' t') = \Gd^+(\r t, \r' t') + \Gd^-(\r t,
  \r' t') .
\eeq
The particle part $\Gd^+$ is defined by
\beq \label{def-Gd+}
  i\Gd^+(\r t, \r' t') = \bpsi{\psi}(\r) e^{-i(H
  -E_{00}) (t-t')} {\psi}^\dagger(\r') \kpsi \,\theta(t-t') .
\eeq
This propagator resembles the standard particle propagator
(\ref{def-G+}) where the electronic Hamiltonian $\He$ appears instead
of the total Hamiltonian $H$ and the electronic ground state energy
$\Vo(\R)$ replaces the molecular ground state energy $E_{00}$. Note
that $H$ now includes the nuclear kinetic energy $\Tn$, which is a
differential operator on the nuclear coordinates $\R$. Thus, also
$\Gd$ is an operator on $\R$. As will be shown explicitly in
Sec.~\ref{sec:exp-opt-pot-Hn}, the dynamical Green's function $\Gd$ may
also be understood as a nonlocal integral operator in the nuclear
coordinates:
\beq
  \bra{\R} \Gd(\r t, \r' t') \ket{\phi} = \int d\R' \Gd(\r \R t, \r'
  \R' t') \, \phi(\R') .
\eeq
For the sake of simple notation, however, we will
not indicate the $\R$-dependence of $\Gd$ explicitly.

The hole part $\Gd^-$ is defined with a modified Hamiltonian because
this will be helpful later for the formulation of the Dyson equation:
\beq \label{def-Gd-}
  i\Gd^-(\r t, \r' t') = \bpsi {\psi}^\dagger(\r') e^{i[H - E_{00} -
  2( \Hn - E_{00})](t-t')}{\psi}(\r) \kpsi\,\theta(t'-t) .
\eeq
As long as we describe scattering or attachment processes, only
$\Gd^+$ has physical significance and the propagation with the
``wrong'' Hamiltonian $H - E_{00} - 2( \Hn - E_{00})$ does not enter
the physics. If, however, we want to study ionisation where the hole part
carries physical significance, we have to modify the definition of the
dynamical Green's function $\Gd$ and allow $\Gd^-$ to propagate with
the correct Hamiltonian $H - E_{00}$. In this case we may modify
$\Gd^+$ in order to obtain a well-behaved Dyson equation. In the
following we will keep to the choice (\ref{def-Gd+}) and
(\ref{def-Gd-}) because we are mainly interested in the scattering
problem in this communication.

Note that the particle part $\Gd^+$ can still be expressed by an
expectation value of a product of Heisenberg field operators:
\beq \label{Gd+Heisenberg}
  i\Gd^+(\r t, \r' t') = e^{-i(\Hn -E_{00}) t} \bpsi \psi(\r t)
  \psi^\dagger(\r t') \kpsi e^{i(\Hn -E_{00}) t'} \theta(t-t') .
\eeq
In contrast to the definition of the purely electronic single-particle
Green's function (\ref{def-G}), the field operators $\psi(\r t)$ and
$\psi^\dagger(\r t')$ now evolve in time with the full molecular
Hamiltonian $H = \Tn + \He$ and thus feel the effect of nuclear motion
in the target. The hole part $\Gd^-$ cannot be expressed in such a
simple way as $\Gd^+$.

We conclude this subsection with the expression the
dynamical Green's function in the frequency domain and orbital
representation: 
\beqs \label{Gd-pq-omega}
  \nonumber
  \Gd_{pq}(\omega) &=& \bpsi a_p \frac{1}{\omega - H +
  E_{00} + i\eta} a^\dagger_q \kpsi \\
  && + \bpsi a^\dagger_q
  \frac{1}{\omega - 2 \Hn + E_{00} + H - i\eta} a_p \kpsi .
\eeqs
This expression is equivalent to the definition of the dynamical
Green's function in the time domain, (\ref{Gd-sum}) to
(\ref{def-Gd-}). The transformation to frequency and orbital
representation is defined as before [see Eq.~(\ref{G-pq-omega})].

\subsection{The inelastic Green's function}

The dynamical Green's function $\Gd$ presents, like the purely
electronic propagator $G$, a matrix (or kernel of an integral
operator) in time and in the coordinates of a single
electron. Moreover, $\Gd$ is an operator in the nuclear
coordinates $\R$. For later use and for comparison with the
(electronically) inelastic theory of reference \cite{cederbaum96}, we
define the {\bf inelastic Green's function} $\Gi^{[m,n]}$ as the
matrix element of $\Gd$ with respect to the nuclear wavefunctions
$\phi_n(\R)$:
\beq \label{def-G-inelastic}
  \Gi^{[m,n]}(\r t, \r' t') = \bra{\phi_m} \Gd(\r t, \r'
  t') \ket{\phi_n} .
\eeq
This inelastic Green's function relates directly to the $S$-matrix of
rotationally or vibrationally inelastic scattering like we will see
later.

The relation to the inelastic Green's function studied by Cederbaum
\cite{cederbaum96} becomes apparent when we evaluate the particle part
of $\Gi^{[m,n]}$. We introduce the symbolic notation $\kket{0k}$ for the
molecular state $\kpsi \phi_k(\R)$ of Eq.~(\ref{mol-eigenstates}),
which describes a (ro-)vibrational excitation with quantum number $k$ in
the electronic ground state. From Eqs.~(\ref{rovib-eigenstates}) and
(\ref{Gd+Heisenberg}) it follows that we can write the particle
part of the inelastic Green's function as
\beq \label{expr-Gd+}
  i\Gi^{[m,n]+}(\r t, \r' t') = \bbra{0m} \psi(\r t)
  \psi^\dagger(\r t') \kket{0n} e^{-i[(E_{0m}-E_{00})t -
  (E_{0n}-E_{00})t']} \theta(t-t') ,
\eeq
where the double bracket notation $\langle \langle \cdot |  \cdot
\rangle \rangle$ indicates integration over both electronic and
nuclear degrees of freedom.
Expression (\ref{expr-Gd+}) is completely analogous to the definition of
Ref.~\cite{cederbaum96}. While in this Reference inelastic scattering
between different electronic excitations of the target is considered
and the nuclear degrees of freedom are neglected, the present study
focuses on inelastic processes with respect to the vibrations and
rotations in the same adiabatic electronic configuration. In contrast
to Ref.~\cite{cederbaum96} where the projectile particle, for instance
a positron, is
distinguishable from the target electrons, our projectile is an
electron and we fully account for
the indistinguishable nature of the projectile and target electrons in
the present approach. For this reason we need the hole part $\Gd^-$,
which, admittedly, takes on a less intuitive form than the particle
part $\Gd^+$ but will prove very useful later for the formulation of
the Dyson equation.

We now want to show that the dynamical Green's function $\Gd$ boils
down to the usual electronic Green's function $G$ when the nuclear
dynamics becomes unimportant. Formally the transition from $\Gd$ to
$G$ is achieved by assuming that the Hamiltonian of nuclear motion
$\Hn = \Tn + \Vo(\R)$ can be replaced by its lowest eigenvalue
$E_{00}$. When we express the nuclear kinetic energy $\Tn$ in the
molecular Hamiltonian $H = \Tn + \He$ by $\Hn - \Vo(\R)$ and replace
$\Hn$ by $E_{00}$, the parts of the dynamical Green's function $\Gd^+$
and $\Gd^-$ in Eqs.~(\ref{def-Gd+}) and (\ref{def-Gd-}) reduce to the
conventional expressions (\ref{def-G+}) and (\ref{def-G-}) for the
electronic Green's function $G^+$ and $G^-$. 
Also in the inelastic formalism the transition to the purely
electronic Green's function can be achieved. The elastic channel
component $\Gi^{[0,0]}$
can be expressed through the purely electronic Green's function under
the assumption that the nuclear kinetic energy $\Tn$ commutes with
the electronic Hamiltonian $\He$:
\beq
  \Gi^{[0,0]}(\r t, \r' t') \stackrel{[\Tn\,,\,\He] \rightarrow
  0}{\longrightarrow} \int 
  d\R\, \phi_0^*(\R)\, G(\r t, \r' t')\, \phi_0(\R) .
\eeq
For the inelastic components we find an expression that reminds us of
the adiabatic-nuclei approximation:
\beq
  \Gi^{[n,m]}(\r t, \r' t') \stackrel{[\Tn\,,\,\He] \rightarrow
  0}{\longrightarrow} \int 
  d\R\, e^{-i(E_{0m} - E_{00})t} \phi_n^*(\R)\, G(\r t, \r' t')\,
  \phi_m(\R) e^{i(E_{0n} - E_{00})t}.
\eeq
The phase factors account for the different reference energies of the
channel states.

\section{Dyson equation and dynamical self energy} \label{sec:dyson}

In this section we will derive Dyson's equation for the dynamical
Green's function and we will show that the dynamical self energy can
easily be expressed as a generalisation of the purely electronic self
energy. Before turning to the dynamical Green's function, however, we
will briefly review the algebraic derivation of Dyson's equation for
the traditional single-particle Green's function. The Dyson equation
for the dynamical Green's function will then follow easily as a
generalisation of this procedure.

\subsection{The Dyson equation for the electronic Green's function}
\label{ssec:trad}

The following derivation of the Dyson equation is close in spirit to
Refs.~\cite{aTarantelli92} and \cite{capuzzi96}. In the present form
it has already been given in Ref.~\cite{brand96}. For convenience we
use the representation of the Green's functions in the frequency
domain and orbital representation in this Chapter as defined by
Eqs.~(\ref{G-pq-omega}) and (\ref{Gd-pq-omega}).

The definition (\ref{G-pq-omega}) of the electronic Green's function
may be written in a more compact form, combining the particle and the
hole propagator. To this end we introduce the following composite
vectors
\beq \label{def-ky}
  \ky{p} = \left({a^\dagger_p \kpsi \atop \bpsi a^\dagger_p } \right)
\eeq
and the matrix
\beq \label{def-Hb}
  \Hb = \left( \begin{array}{cc} \He - \Vo(\R) & 0 \\
                               0             & \Vo(\R) - \He
       \end{array} \right) .
\eeq

We can rewrite Eq.~(\ref{G-pq-omega}) by formally expressing the
single-particle Green's function as a matrix element of the resolvent
of $\Hb$ with respect to the states $\ky{p}$:
\beq \label{G-compact}
  G_{pq}(\omega) = \by{p} \frac{1}{\omega - \Hb} \ky{q} .
\eeq
Here, we omitted the $i\eta$ terms to avoid clumsy notation. The
time-ordering controlled by these terms is usually not important but
may become relevant in time-dependent formulations of scattering
theory (see Sec.~\ref{scat-deriv}).

The scalar product of the composite vectors $\ky{p}$ now includes the
matrix-vector product where bras always move to the left and kets move
to the right.  With respect to this canonical scalar product, the
vectors $\ky{p}$ fulfil the following orthonormality relation:
\beqs
  \nonumber
  \bracket{Y_p}{Y_q} &=& \bpsi a_p a^\dagger_q \kpsi +
  \bpsi a^\dagger_q   a_p \kpsi 
  = \delta_{pq} .
\eeqs

When the $N$-electron wavefunction $\kpsi$ is approximated by a Slater
determinant, the vector $\ky{p}$ contains either a single-hole state,
if $p$ refers to an unoccupied orbital, or a
single-particle state, if not. In the more general case
of a correlated wavefunction $\kpsi$, the vectors $\ky{p}$ mix $N-1$
and $N+1$-electron states.  Note that in either case and for arbitrary
single-particle indices $p$, the vectors $\ky{p}$ may be seen as
orthogonal states that span a linear space with the same dimension as
(or isomorphous to) the single-particle Hilbert space. We will call
the space spanned by the vectors $\{\ky{p}\}_p$ the
{\bf primary space}. Since the primary states $\ky{p}$ and their linear
combinations are not, in general, exact eigenstates of the electronic
Hamiltonian $\He$, they couple to the higher excitations like
two-particle--one-hole excitations, one-particle--two-hole excitations,
etc. The space of higher excitations will be called the {\bf secondary
space} \cite{note-Y-space}.
The concept of single-hole, two-particle--one-hole excitations, etc.~is,
of course, only adequate when the target state $\kpsi$ is dominated by
a single configuration. In general, $\kpsi$ will be represented by a
correlated wavefunction and thus the primary and the secondary
space cannot be easily expressed by single-configuration basis
states. For an explicit construction of a basis for the secondary
space that allows for convenient approximations to the Green's
function or the self energy in the framework of the so-called
intermediate state 
representations see Refs.~\cite{aTarantelli92,mertins96:II}.
In the following we assume that we have a basis of the composite space
$\{\ket{Q_J}\}_J$ that consists of the primary states $\{\ky{p}\}_p$
augmented by any suitable basis for the secondary space. The particular
choice of the basis for the secondary space does not matter.

We may now proceed to derive the Dyson equation and define the self
energy setting out from Eq.~(\ref{G-compact}). The basis
$\{\ket{Q_J}\}_J$ of the composite space defines a basis set
representation $\mat{\Hb}$ of the matrix Hamiltonian $\Hb$:
\beq \label{def-mat-Hb}
  [\mat{\Hb}]_{IJ} = \bra{Q_I} \Hb \ket{Q_J} .
\eeq
By virtue of the subdivision of the basis set into two parts, the
matrix $\mat{\Hb}$ is structured into blocks:
\beq \label{struc-mat-Hb}
  \mat{\Hb} = \left( \begin{array}{cc} 
                       \mat{\Hb}_{aa} & \mat{\Hb}_{ab}  \\
                       \mat{\Hb}_{ba} & \mat{\Hb}_{bb}
              \end{array} \right) .
\eeq
The block index $a$ refers to primary states $\{\ky{p}\}_p$ and $b$ to
the rest of the basis $\{\ket{Q_J}\}_J$. The upper left block of this
matrix is readily evaluated using Eqs.~(\ref{def-ky}) and (\ref{def-Hb}):
\beq \label{ev-Hb-aa}
  \left[ \mat{\Hb}_{aa} \right]_{pq} = \by{p} \Hb \ky{q}
  = \bpsi \{ a_p\, , \,[ \Hb\, , \,a^\dagger_q ] \} \kpsi .
\eeq

It is easily seen, now, that the electronic Green's function of
Eq.~(\ref{G-compact}) can be understood as the upper left block of an
inverse matrix:
\beq \label{inv-matr-aa}
   \mat{G}(\omega) = \left( \frac{1}{\omega \mat{1} - \mat{\Hb}}
   \right)_{aa}  .
\eeq
The proof follows by using the completeness of the basis
$\{\ket{Q_J}\}_J$ and the orthonormality of the primary states
$\{\ky{p}\}_p$. By simple matrix partitioning, the inverse matrix of
the Green's function $\mat{G}(\omega)$, can be expressed by
\beq \label{inv-Gf}
  \mat{G}(\omega)^{-1} = \omega \mat{1} - \mat{\Hb}_{aa} -
  \mat{\Hb}_{ab} \frac{1}{\omega \mat{1} - \mat{\Hb}_{bb}}
  \mat{\Hb}_{ba} .
\eeq

In order to define the self energy and to make contact to the usual
form of Dyson's equation, we introduce a perturbation theoretic
separation of the electronic Hamiltonian $\He$ into a zeroth order, an
interaction part, and the internuclear repulsion term $U_{\text{N}}(\R)$,
which does not act as an operator on the electronic coordinates:
\beq \label{part-He}
  \He = \Hez + \Hei + U_{\text{N}}(\R) .
\eeq
The zeroth order part $\Hez$ is a one-particle operator of the form
\beq \label{def-Hez}
  \Hez = \sum_{ij} \varepsilon_{ij} \, a_i^\dagger a_j .
\eeq
A convenient choice may be realized by the electronic kinetic energy
or the Hartree-Fock operator. We do not demand the operator $\Hez$ to
be diagonal in the given orbital basis but a diagonalising choice of
the single-particle basis is always possible. E.~g.~if $\Hez$ is
chosen to represent the kinetic energy, the matrix $\mat{\varepsilon}$
with the matrix elements $\left[ \mat{\varepsilon}\right]_{ij} =
\varepsilon_{ij}$ becomes diagonal in the momentum representation. In
atomic units, the diagonal element is then simply given by
$\varepsilon_p = {p^2}/{2}$. In practical applications,
$\Hez$ is often chosen the Hartree-Fock operator.

On the assumption that the electronic ground-state wavefunction
$\kpsi$ and energy $\Vo(\R)$ both possess well-behaved perturbation
expansions,
the zeroth order of the matrix $\mat{\Hb}_{aa}$ can be easily
evaluated from Eq.~(\ref{ev-Hb-aa}) to give the matrix of the
zeroth-order orbital energies:
\beq \label{ev-Hb-aa-zo}
  \mat{\Hb}_{aa}^{(0)} = \mat{\varepsilon} .
\eeq
It is also easy to see that the zeroth orders of the off-diagonal
blocks $\mat{\Hb}_{ab}$ and $\mat{\Hb}_{ba}$ vanish and hence the
zeroth-order electronic Green's function $G^{(0)}$ is found from
Eq.~(\ref{inv-Gf}) to yield
\beq
  \mat{G}^{(0)}(\omega) = \frac{1}{\omega \mat{1} - \mat{\varepsilon}} .
\eeq
The electronic self energy is defined by
\beq \label{elec-self-energy}
  \mat{\Sigma}(\omega) = \mat{\Hb}_{aa} - \mat{\varepsilon} +
  \mat{\Hb}_{ab} \frac{1}{\omega \mat{1} - \mat{\Hb}_{bb}}
  \mat{\Hb}_{ba} 
\eeq
and Eq.~(\ref{inv-Gf}) remains
\beq
  \mat{G}(\omega) = \frac{1}{\omega \mat{1} - \mat{\varepsilon}-
  \mat{\Sigma}(\omega)} .
\eeq
This is equivalent to the common form of Dyson's equation:
\beq
  \mat{G}(\omega) = \mat{G}^{(0)}(\omega) +  \mat{G}^{(0)}(\omega)
  \mat{\Sigma}(\omega)\mat{G}(\omega) .
\eeq

\subsection{The Dyson equation for the dynamical Green's function}
\label{sec:dyn-gf} 

We will show in the following that a Dyson equation for the dynamical
Green's function $\Gd$ can be derived in an analogous manner, using
the concepts of the previous section. We will discuss in particular
the dependence on the nuclear coordinates, which becomes important
because of the introduction of the nuclear kinetic energy operator.

Just like the electronic Green's function, the dynamical Green's function
$\Gd$ of Eq.~(\ref{Gd-pq-omega}) can be written in a compact form
combining the ionisation and attachment parts. Using the composite
states $\ky{p}$ of (\ref{def-ky}) and the electronic matrix
Hamiltonian $\Hb$ of (\ref{def-Hb}), we can rewrite the expression
(\ref{Gd-pq-omega}) for the dynamical Green's function to give
\beq \label{Gd-compact}
  \Gd_{pq}(\omega) = \by{p} \frac{1}{\omega - \Hb - \Hn + E_{00}}
  \ky{q} . 
\eeq
We again dropped the $i\eta$ terms here for brevity. Also the $2
\times 2$ unit matrix that allows $\omega$, $\Hn$, and $E_{00}$ to be
applied to the two-component vector $\ky{p}$ is omitted. The particularly
simple and compact form of this expression can be seen as a
preliminary justification for the particular choice of $\Gd^-$ in the
definition (\ref{def-Gd-}). Dyson's equation can now easily be derived
in an analogous way to the last section.

Comparing with the expression (\ref{G-compact}) for the electronic
Green's function, we see that only the term $-\Hn + E_{00}$ is
additionally present in the dynamical Green's function of
Eq.~(\ref{Gd-compact}). The molecular ground state energy $E_{00}$ is
just a constant and defines the zero point of the $\omega$ scale. The
``nuclear'' Hamiltonian $\Hn = \Tn + \Vo(\R)$ introduces the nuclear
kinetic energy operator $\Tn$, which is a differential operator on the
nuclear coordinates $\R$. It is therefore useful to consider the
$\R$-dependence of the quantities used in the derivation of the Dyson
equation.

The matrix operator $\Hb$ contains the electronic Hamiltonian $\He$
and the ground state energy $\Vo(\R)$. Both quantities depend on the
nuclear coordinates $\R$ and thus care has to be taken because $\Hb$
and $\Hn$ do not commute. For the moment we want to assume that the
electronic orbital basis that defines the electronic creation
operators $a^\dagger_p$ is independent of the nuclear coordinates
$\R$. This is the case, e.~g.~for the momentum or position space
representation. This restriction is just convenient for the derivation
of the Dyson equation but not essential and we will see later
(Sec.~\ref{sec:opt-pot-proj}) that it can be lifted as long as the
chosen single-particle basis for the electrons is complete. The $\R$
dependence of the composite states $\ky{p}$ of Eq.~(\ref{def-ky}) now
derives entirely from the electronic ground state $\kpsi$. Like we
already mentioned in Sec.~\ref{ssec:def-dyn}, we assume that the
Born-Oppenheimer approximation is valid for the electronic ground
state and thus the nuclear kinetic energy $\Tn$ commutes with
$\kpsi$. Consequently we may assume that the nuclear Hamiltonian $\Hn$
commutes with the composite states $\ky{p}$.  We also assume that the
basis states $\ket{Q_J}$ of the secondary space are conveniently
chosen, such that the commutator with $\Tn$ or $\Hn$ can be
neglected. An appropriate choice is always possible
\cite{note-projection}.
The nuclear Hamiltonian $\Hn$ can consequently be pulled out of any
matrix element involving the basis states of the composite electronic
space and thus the matrix representation of $\Hn$ in this basis is
proportional to the unit matrix:
\beq
  \bra{Q_I} \Hn \ket{Q_J} = \delta_{IJ} \Hn .
\eeq
In analogy to Eq.~(\ref{inv-matr-aa}) we can understand the dynamical
Green's function $\Gd$ as the upper left block of an inverse matrix:
\beq
  \mat{\Gd}(\omega) = \left( \frac{1}{(\omega -\Hn + E_{00})\mat{1} -
   \mat{\Hb}} \right)_{aa}  .
\eeq
The elements of this matrix are now operators acting on the nuclear
coordinates. Matrix partitioning can be applied and is completely
analogous to the case of the electronic Green's function. 

Owing to the partitioning (\ref{part-He}) of the electronic
Hamiltonian $\He$ into a zeroth order and interaction part, we find
for the zeroth order dynamical Green's function
\beq
  \mat{\Gd}^{(0)}(\omega) = \frac{1}{(\omega -\Hn + E_{00})\mat{1}
  -\mat{\varepsilon}} ,
\eeq
where $\mat{\varepsilon}$ is again the matrix of zeroth-order
single-electron energies. In the simple case where the electronic
kinetic energy is chosen as the zeroth-order electronic Hamiltonian
$\Hez$, the matrix $\mat{\varepsilon}$ is diagonal in the momentum
representation and independent of the nuclear coordinates $\R$ as
mentioned in the Sec.~\ref{ssec:trad}. The zeroth order dynamical
Green's function $\Gd^{(0)}$ now is also diagonal and describes the
simultaneous motion of a free electron and the vibrations (or
rotations) of the isolated target molecule.

In the more general case where the matrix $\mat{\varepsilon}$ depends
on $\R$ (if, e.~g.~$\Hez$ is chosen the $\R$-dependent Hartree-Fock
operator), the zeroth-order propagator $\Gd^{(0)}$ may be understood
to describe the coupled motion of a vibrating (or rotating) molecule
in the electronic ground state and a single electron that moves under
the influence of $\R$-dependent mean fields. Since the zeroth-order
electronic Hamiltonian $\Hez$ is a one-particle operator, the motion
of the scattering electron is decoupled from the target electrons. The
effective Hamiltonian for the coupled projectile and nuclear motion is
thus independent of the scattering energy and simply given by $\Hn +
\Hez - E_{00}$. The zeroth-order Green's function $\Gd^{(0)}$ is
the resolvent of this effective Hamiltonian.

For the full dynamical Green's function we find
\beq
  \mat{\Gd}(\omega) = \frac{1}{(\omega -\Hn + E_{00})\mat{1}
  -\mat{\varepsilon} - \mat{\Sd}(\omega)} ,
\eeq
where the dynamical self energy $\Sd$ is defined by
\beq \label{self-energy}
  \mat{\Sd}(\omega) = \mat{\Hb}_{aa} - \mat{\varepsilon} +
  \mat{\Hb}_{ab} \frac{1}{(\omega -\Hn + E_{00}) \mat{1} - \mat{\Hb}_{bb}}
  \mat{\Hb}_{ba} .
\eeq
In comparison to the zeroth order case, the dynamical self energy
$\Sd(\omega)$ now accounts for the many-body nature of the molecular
target. We will demonstrate in the next section that $\Sd(\omega)$ indeed is
an optical potential for (electronically elastic) electron-molecule
scattering. It is remarkable that the nuclear Hamiltonian $\Hn$
appears in the self energy $\Sd(\omega)$ together with the $\omega$
dependence and thus introduces derivatives with respect to the nuclear
coordinates. Like we shall see in detail later, the dynamical self
energy thus becomes a non-local operator not only in the electronic
but also in the nuclear coordinates. In Sec.~\ref{sec:approximate-Tn}
we will discuss the meaning of the nuclear Hamiltonian appearing in the
dynamical self energy and possible approximations.

In a very formal manner, $\Sd(\omega)$ can be expressed by
the usual electronic self energy $\Sigma(\omega)$:
\beq \label{Sd-by-Sigma}
  \mat{\Sd}(\omega) = \mat{\Sigma}(\omega -\Hn + E_{00}).
\eeq

Before we turn to the discussion of the scattering $S$-matrix and the
optical potential in the next section, we would like to present the
Dyson equation in terms of the inelastic Green's function
$\Gi^{[m,n]}$ of Eq.~(\ref{def-G-inelastic}). Carrying on the analogy
to the inelastic formalism of Ref.~\cite{cederbaum96} will help us in
the next section to identify the dynamical self energy with the
optical potential. The inelastic formalism also yields a slightly
different view of the Dyson equation as the introduction of
nuclear eigenfunctions $\phi_n(\R)$ formally leads to an equal
treatment of the electronic and nuclear degrees of freedom, at the
cost of having to introduce another sort of matrix indices.

Instead of using the purely electronic basis $\{ \ket{Q_J} \}_J$, we
introduce the product basis $\{ \ket{Q_J} \otimes \ket{\phi_n}
\}_{J,n}$, where $\ket{\phi_n}$ is the abstract notation for the
nuclear wavefunction $\phi_n(\R)$. The expression
(\ref{def-G-inelastic}) for the inelastic Green's function in the
frequency domain and orbital representation then reads
\beq
  \Gi^{[m,n]}(\omega) = \bra{\phi_m} \by{p} \frac{1}{\omega - \Hb - \Hn
  + E_{00}}  \ky{q} \ket{\phi_n} . 
\eeq
The bracket of the nuclear states $\bra{\phi_m} \ldots \ket{\phi_n}$
implies integration over the nuclear coordinates. The derivation
of the Dyson equation can now be redone using the matrix representation
defined by the product basis. In the following we denote matrices in
the labels of the nuclear states $n,m$ by boldface letters and
matrices in the electronic state labels $I,J,p,q$ with double
underbars. The zeroth order inelastic Green's function becomes
\beq
  \mat{\bbox \Gi}^{(0)}(\omega) = \frac{1}{(\omega + E_{00}) \mat{\bbox 1} -
  {\bHn} \mat{1} - \mat{\bbox \varepsilon}} . 
\eeq
The matrix of the nuclear Hamiltonian $\Hn$ is diagonal due to the
particular choice of the nuclear wavefunction basis:
\beq
  [{\bHn}]^{[m,n]} = \delta_{mn} E_{0n} .
\eeq
The full inelastic Green's function reads
\beq
  \mat{\bbox \Gi}(\omega) = \frac{1}{(\omega + E_{00}) \mat{\bbox 1} -
  {\bHn} \mat{1} - \mat{\bbox \varepsilon} - \mat{\bbox \Si}(\omega)} , 
\eeq
where the inelastic self energy $\mat{\bbox \Si}(\omega)$ is defined by
\beq
  \mat{\bbox \Si}(\omega) = \mat{\bHb}_{aa} - \mat{\bbox{\varepsilon}} +
  \mat{\bHb}_{ab} \frac{1}{(\omega + E_{00}) \mat{\bbox 1} -{\bHn}
  \mat{1} - \mat{\bHb}_{bb}} \mat{\bHb}_{ba} .
\eeq
Of course, matrix inversion and multiplication now have to include
both sorts of matrix indices. The Dyson equation for the inelastic
Green's function can be written as
\beq \label{Dys-eq-inelast}
  \mat{\bbox \Gi}(\omega) =   \mat{\bbox \Gi}^{(0)}(\omega) + \mat{\bbox
  G}^{(0)}(\omega) \,\mat{\bbox \Si}(\omega)\, \mat{\bbox \Gi}(\omega) .
\eeq
Note that the inelastic self energy $\Si$ is merely another
representation of the dynamical self energy $\Sd$. The relation is
given by
\beq
  \Si^{[m,n]}_{pq}(\omega) = \bra{\phi_m}\, \Sd_{pq}(\omega)\,
  \ket{\phi_n}  .
\eeq


\section{Application to electron-molecule scattering} \label{sec:scat}

In this chapter we discuss the relation of the dynamical Green's
function to the process of electron-molecule scattering. We introduce
the $S$- and $T$-matrices of electronically elastic but vibrationally
or rotationally inelastic scattering and show how they can be
calculated using the dynamical Green's function formalism developed in
the last sections. We present an effective Schr\"odinger equation and
discuss the role of the dynamical self energy as an optical potential
for the scattering process. An direct derivation of the effective
Schr\"odinger equation is
given and possible approximations to the exact equations are
discussed.

\subsection{The $S$- and $T$-matrices of inelastic scattering}
\label{scat-deriv}

The relation of the dynamical Green's function to the $S$- and
$T$-matrices of scattering theory is very similar to elastic
electronic scattering off atoms or rigid molecules
\cite{csanak71,ring80} and completely analogous to the electronically
inelastic case \cite{cederbaum96}. We therefore sketch the derivation
only very briefly.

The $S$-matrix for the processes we consider reads
\beq \label{s-matrix}
  S(p'm \leftarrow pn) = 
  \bbracket{\Psi^{m-}_{p'}}{\Psi^{n+}_{p}} , 
\eeq
where the stationary scattering states $\kket{\Psi^{m\pm}_{p}}$ with
incoming ($-$) or outgoing ($+$) boundary conditions are defined
by
\beq \label{scat-state}
  \kket{\Psi^{n\pm}_{p}} = \lim_{t \rightarrow \mp \infty}
  e^{-i(\varepsilon_p +E_{0n} - H) t} a^\dagger_p \kpsi\, \ket{\phi_n} ,
\eeq
and the double bracket notation again implies integration over
electronic and nuclear degrees of freedom.
The asymptotic states thus feature a free electron with momentum $p$
and energy $\varepsilon_p = p^2/2$ (in atomic units) and an
$N$-electron molecular state $\kpsi \ket{\phi_n}$ with the $n$th
excited vibration or rotation in the electronic ground state with
energy $E_{0n}$. For definiteness we choose the electronic kinetic
energy as the zeroth-order electronic Hamiltonian $\Hez$ in this
section, describing the motion of free electrons. Note that the scalar
product of Eq.~(\ref{s-matrix}) implies integration over all
coordinates of the $N+1$ electrons in addition to the 
integration over the nuclear coordinates $\R$.

The $S$-matrix can be expressed by the particle part of the
inelastic Green's function of Eq.~(\ref{def-G-inelastic}) in the
momentum representation as follows from the definition
(\ref{def-Gd+}) and Eqs.~(\ref{s-matrix}) and (\ref{scat-state}):
\beq
  S(p'm \leftarrow pn) = \lim_{t \rightarrow + \infty \atop t'
  \rightarrow -\infty} e^{i(\varepsilon_{p'} + E_{0m} - E_{00}) t}
  \,\Gi^{[m,n]+}_{p'p}(t,t')\, e^{-i(\varepsilon_p + E_{0n} - E_{00})
  t'} . 
\eeq
The particle part $\Gi^{[m,n]+}$ may be replaced here by the full Green's
function $\Gi^{[m,n]}$ for two reasons:
\begin{itemize}
\item The hole part $\Gi^{[m,n]-}$ does not contribute because the
 $S$-matrix only contains information about the asymptotic region of
 scattering where $\Gi^{[m,n]-}$ vanishes because $a_p \kpsi$
 vanishes. For the effective single-particle scattering equations we
 are going to derive in the following, this means that ensuring the
 correct outgoing boundary condition avoids contamination of the
 wavefunction by unphysical contributions.
\item The hole part $\Gi^{[m,n]-}$ does not contribute in the
 considered time limit because of the theta functions that appear
 explicitly in Eqs.~(\ref{def-Gd+}) and (\ref{def-Gd-}). Therefore
 the particle part may be replaced by the full Green's function when the
 time ordering is treated properly. 
\end{itemize}
We now introduce the {\bf inelastic improper self energy} $\mat{\bbox
\Ti}(\omega)$  in analogy to the common improper or reducible self
energy \cite{fetter71,ring80} by rewriting Eq.~(\ref{Dys-eq-inelast})
as usual by 
\beqs
  \mat{\bbox \Gi}(\omega) &=& \mat{\bbox \Gi}^{(0)}(\omega) + \mat{\bbox
  \Gi}^{(0)}(\omega) \, \mat{\bbox \Ti}(\omega) \, \mat{\bbox
  \Gi}^{(0)}(\omega) , \\ \label{T-matrix-by-A}
  \mat{\bbox \Ti}(\omega) &=& \mat{\bbox \Si}(\omega) + \mat{\bbox
  \Si}(\omega)\, \mat{\bbox \Gi}^{(0)}(\omega) \, \mat{\bbox \Ti}(\omega) .
\eeqs
Following the derivations of Refs.~\cite{ring80,cederbaum96} we can
express the scattering $S$-matrix by the improper self energy:
\beq
  S(p'm \leftarrow pn) = \delta_{nm} \delta_{pp'} -2\pi i \,
  \Ti^{[m,n]}_{p'p}(\varepsilon_p + E_{0n})\, \delta(\varepsilon_{p'} +
  E_{0m} - \varepsilon_p - E_{0n}) .
\eeq
The inelastic improper self energy $\Ti^{[m,n]}$ at the scattering energy
$\varepsilon_p + E_{0n}$ can thus be identified with the on-shell
$T$-matrix of scattering theory. Following standard scattering theory,
we see from Eq.~(\ref{T-matrix-by-A}) that the inelastic self energy
${\Si}^{[m,n]}(\omega)$ (or equivalently the dynamical self
energy $\Sd(\omega)$) presents the (optical)
potential for the scattering process. $\Si^{[m,n]}$ is the optical
potential for multichannel scattering where the channels are defined
by the eigenstates $\ket{\phi_n}$ of the nuclear motion.

The relation of the inelastic improper self energy $\Ti$ to the
improper self energy $T$ of the usual (electronic) single-particle
Green's function (\cite{fetter71,ring80}, see also \cite{kutzelnigg89}
for explicit expressions that can be useful in the present context) can be
described, in analogy to Eq.~(\ref{Sd-by-Sigma}), by
\beq
  \mat{\Ti}^{[m,n]}(\omega) = \bra{\phi_m} \mat{T}(\omega -\Hn +
  E_{00}) \ket{\phi_n} .
\eeq
Note that this formal expression in fact expresses a rather
complicated relation of the purely electronic improper self energy
$\mat{T}(\omega)$ and the inelastic scattering $T$-matrix because the
nuclear kinetic energy in $\Hn$ formally appears as an energy variable
of $\mat{T}(\omega)$, which has itself a complicated dependence on the
nuclear coordinates $\R$. If, however, $\Tn$ is omitted in this
expression replacing $\Hn$ by $\Vo(R)$, we can recover the well-known
adiabatic-nuclei approximation (an) of electron-molecule scattering:
\beq
  \Ti^{[m,n] {\text{(an)}}}_{p'p}(\omega) = \bra{\phi_m}
  {T}_{p'p}(\omega -\Vo(\R) + E_{00}) \ket{\phi_n} .
\eeq

\subsection{An effective Schr\"odinger equation} \label{sec:eff-schr-eq}

In the preceding section we identified the inelastic self energy
$\Si^{[m,n]}$ with the optical potential for multi-channel
scattering. Consequently, effective one-particle equations of the
Lippmann-Schwinger or Schr\"odinger type can be derived. In the
multichannel picture one obtains sets of coupled equations, one for
each combination of initial and final channel. Instead of rederiving
these equations, which can be found in Ref.~\cite{cederbaum96} for the
analogous case of electronically inelastic scattering, we
will discuss directly the time-independent Schr\"odinger equation in the
position-space representation.

Any effective one-particle equation must involve explicitly the
coordinates of the scattering electron and the labels of the nuclear
states or, equivalently, the nuclear coordinates $\R$. Instead of the
full scattering wavefunction 
\beqs \label{tot-scat-wf}
  \nonumber
  \ket{\Psi^{n+}_{p}({\R})} &=& \brakket{\R}{\Psi^{n+}_{p}} \\
  & = & \lim_{t \rightarrow -\infty}
  e^{-i(\varepsilon_p +E_{0n} - H) t} a^\dagger_p \kpsi\,{\phi_n(\R)},
\eeqs
which is the nuclear-coordinate representation of the scattering
wavefunction of Eq.~(\ref{scat-state}) and presents a
wavefunction in the nuclear coordinates and an abstract ket in the
$N+1$-particle space, we introduce the 
effective or {\bf optical wavefunction} $f(\r,\R)$ by
\beq \label{def-scat-ampl}
  f^{[n]+}_p(\r,\R) = \bpsi \psi(\r) \,\ket{\Psi^{n+}_{p}({\R})} .
\eeq
This function can be calculated from the dynamical Green's function
$\Gd$ of Eq.~(\ref{Gd-sum}) by
\beq
  f^{[n]+}_p(\r,\R) = \lim_{t' \to \infty} \int d\r'\, \bra{\R}
  \Gd(\r,0;\r',t') \ket{\phi_n}
  \, \varphi_p(\r') e^{-i(\varepsilon_p +E_{0n} - E_{00})} ,
\eeq
where $\varphi_p(\r)$ is a plane wave. The relation to the scattering
$S$-matrix is given by \cite{csanak71,cederbaum96}
\beq
  S(p'm \leftarrow pn) = \int d\R \, \int d\r\,
  \phi^*_m(\R)\, \varphi^*_{p'}(\r)\, f^{[n]+}_p(\r,\R).
\eeq
An effective Schr\"odinger equation for the considered scattering
problem reads 
\beq \label{eff-schroed-eq}
  \left[ \Hez + \Hn + {\Sd}(E)  - E_{00}\right]  f^{[n]+}_p(\r,\R)
  = E \, f^{[n]+}_p(\r,\R) ,
\eeq
as follows by analogy from Refs.~\cite{cederbaum96,csanak71,ring80}. A
self-contained derivation of this equation using projection operators
will also be given in Sec.~\ref{sec:opt-pot-proj}.  In the effective
Schr\"odinger equation (\ref{eff-schroed-eq}), $E$ denotes the total
energy with respect to the zero point defined at the molecular
ground-state energy $E_{00}$. This zero point is related to a
situation where the scattering electron is so far away that it does
not feel any forces exerted by the target, which is in its ground
state, and the particle as well as the projectile are at rest. The
zeroth-order electronic Hamiltonian $\Hez$ is an operator acting only
on the coordinates $\r$ of the scattering electron. According to the
choice of Sec.~\ref{scat-deriv}, it consists of the electronic kinetic
energy and describes the motion of a free electron.
Eq.~(\ref{eff-schroed-eq}), however, is also valid if the Hartree-Fock
operator is chosen as the zeroth order. In this case, $\Hez$ contains
also the static-exchange potential of the target molecule.  The
nuclear Hamiltonian $\Hn$ acts only on the nuclear coordinates $\R$
and describes the nuclear motion of the target molecule in its
electronic-ground-state potential $\Vo(\R)$.  The dynamical self
energy ${\Sd}(E)$ of Eqs.~(\ref{self-energy}) and (\ref{Sd-by-Sigma})
appears here in the coordinate-space representation and takes account
of the complex many-body nature of the molecular target. It acts as a
nonlocal (integral) operator in the electron coordinates $\r$ and also
in the nuclear coordinates $\R$:
\beq
  {\Sd}(E)\, f(\r,\R) = \int d\R'\,\int d\r'\, \Sd(\r,\R,\r',\R',E)\,
  f(\r',\R'). 
\eeq

Introducing the operator
\beq
  {\cal{L}}(E) = \Hez + \Hn + {\Sd}(E)  - E_{00} ,
\eeq
which can be seen as an analogue of the Layzer operator
\cite{layzer63} of the usual single-particle Green's function, the
effective Schr\"odinger equation (\ref{eff-schroed-eq}) can be written
as
\beq \label{schr-eq-Layzer}
  {\cal{L}}(E) f(\r,\R) = E\, f(\r,\R) .
\eeq
This is formally a pseudo-eigenvalue equation because the eigenvalue
$E$ also determines the operator ${\cal{L}}(E)$. In the scattering
regime ($E>0$) there is, of course, a solution for every $E$ and the
problem of solving Eq.~(\ref{schr-eq-Layzer}) is to find a
wavefunction $f(\r,\R)$ with the correct boundary conditions such that
the equation is fulfilled at a given scattering energy $E$. We usually
identify the dynamical self energy $\Sd(E)$ with the dynamical optical
potential although $\Sd(E)$ is not the only non-kinetic-energy
component of ${\cal{L}}(E)$.

As it stands, the effective Schr\"odinger equation
(\ref{eff-schroed-eq}) or (\ref{schr-eq-Layzer}) is an exact equation
but, of course, approximations have to be introduced in most cases to
determine the integro-differential operator ${\cal{L}}(E)$. We will
discuss some possibilities in Sec.~\ref{sec:approximate-Tn}. The best
strategy to solve Eq.~(\ref{eff-schroed-eq}) then certainly depends on
the level of approximation introduced in the determination of
${\cal{L}}(E)$ and on the nature of the problem to be solved. A
variety of methods is available. E.~g.~close-coupling
\cite{lane80,morrison95} or the other methods mentioned in
Sec.~\ref{sec:intro} can be used to solve Eq.~(\ref{eff-schroed-eq})
directly for the scattering wavefunctions. Resonances may also be
calculated using bound-state techniques by employing the
complex-rotation method
\cite{aguilar71,reinhardt82}, complex absorbing potentials
\cite{jolicard85,riss93,sommerfeld98-liste}, or the stabilisation
method
\cite{mandelshtam94}, etc. If necessary, the eigenvalues $E$ may be found
by iteration. Any other standard method for solving scattering problems
will also do, since the corresponding Lippmann-Schwinger or effective
time-dependent Schr\"odinger equation may also be used as starting
points.


\subsection{A direct derivation of the scattering equations}
\label{sec:opt-pot-proj} 

We will now show how to derive the effective Schr\"odinger equation
(\ref{eff-schroed-eq}) directly from the Schr\"odinger equation for
the full scattering problem, which comprises the motion of $N+1$
electrons and the nuclear motion. This derivation is fully equivalent
to the derivation sketched in Sec.~\ref{scat-deriv}. Although the
dynamical Green's function and its Dyson equation do not appear
explicitly, the direct derivation, which uses projection operators in a
composite Hilbert space with particle and hole states, is firmly based
on Green's function theory and has to be seen as the result of recent
developments in this area \cite{aTarantelli92,capuzzi96}. 
While traditional many-body Green's function theory supplies powerful
methods to compute the dynamical optical potential approximatively (see
Sec.~\ref{sec:approximate-Tn}), the projection operator formulation of Dyson's
equation \cite{capuzzi96} yields a
complementary way of reflecting on the Green's function approach and
its relation to Feshbach projection.
Reconsidering in this section the algebra
of Sec.~\ref{sec:dyson} will help us to further clarify the role of an
$\R$-dependence in the electronic single-particle basis and in the
electronic ground-state $\kpsi$. We will also have an opportunity
to discuss the role of unphysical contaminations of the optical
wavefunction. 

We start from the time-independent Schr\"odinger equation for the full
scattering problem of $N+1$ electrons and the atomic nuclei:
\beq \label{schr-eq}
  (H-E_{00}) \kpt = E \kpt.
\eeq
The total wavefunction $\kpt$ is an arbitrary solution of the Schr\"odinger
equation like, e.~g., the scattering wavefunction $\ket{\Psi^{n +}_{p}({\R})}$
of Eq.~(\ref{tot-scat-wf}). The eigenvalue $E$ measures the energy of
the scattering system relative to the molecular ground state energy
$E_{00}$.
Following Eq.~(\ref{split-ham}), we split the total Hamiltonian $H$
into the electronic part $\He$ and the nuclear kinetic energy
$\Tn$. Adding and subtracting the ground-state potential $\Vo(\R)$ and
introducing the nuclear Hamiltonian $\Hn = \Tn + \Vo(\R)$ of
Eq.~(\ref{def-Hn}), we may write
\beq \label{umf1}
  [\He - \Vo(\R) + \Hn - E_{00}] \kpt = E \kpt.
\eeq
By introducing the $2\times 2$ matrix
operator $\Hb$ of Eq.~(\ref{def-Hb}) we may rewrite Eq.~(\ref{umf1})
into the two-component equation
\beq \label{umf2}
  \left[ \Hb + (\Hn - E_{00})\mat{1}_2 \right] \left( \kpt \atop 0
  \right) = E \left( \kpt \atop 0 \right),
\eeq
where $\mat{1}_2$ is the $2\times 2$ unit matrix. Since the second
component of the vector $\left( \kpt ,\,0 \right)^t$ is set to zero,
Eq.~(\ref{umf2}) is completely equivalent to the initial Schr\"odinger
equation (\ref{schr-eq}). However, the vector $\left( \kpt ,\,0 \right)^t$
may also be seen as an element of the composite ${\sf Y}$-space
of Note \cite{note-Y-space}. We now
introduce the projection operator
\beq
  P = \sum_q \ky{q}\by{q} ,
\eeq
projecting onto the primary space, which is spanned by the vectors $\{
\ky{p}\}$ of Eq.~(\ref{def-ky}). By definition, the projector $P$
primarily acts on the electronic degrees of freedom. However, we also
have to consider the dependence on the nuclear coordinates $\R$
because the nuclear kinetic energy $\Tn$ contained in the nuclear
Hamiltonian $\Hn$ introduces derivatives with respect to $\R$ into
Eq.~(\ref{umf2}). The nuclear-coordinate dependence of $P$ derives
directly from the electronic ground state $\kpsi$, as can be seen from
the definition (\ref{def-ky}) of $\ky{p}$ when we assume that the
basis of single-particle orbitals $\varphi_p(\r)$ is independent of
$\R$. Since the primary space and consequently the projector $P$ are
invariant under a change of the single-particle basis, this also holds
true when $\R$-dependent orbitals like, e.~g., Hartree-Fock orbitals
are chosen, as long as they from a complete basis of the
single-particle space. Consequently a formulation employing the projection
operators $P$ and $Q = 1 - P$ remains independent of the choice of the
single-particle basis and thus an $\R$-dependence of the orbitals
does not matter.

We already discussed in Sec.~\ref{sec:dyn-gf} that we want to adopt
the Born-Oppenheimer approximation for the electronic ground state
$\kpsi$. In particular we assume the $\R$-dependence of the electronic
ground state $\kpsi$ to be weak. As a consequence, the projection
operator $P$ commutes with the nuclear kinetic energy $\Tn$ and
therefore also with $\Hn$:
\beq \label{approx-comm}
  [P\,,\,\Tn] = [P\,,\,\Hn] = 0 .
\eeq
We want to stress once more that we assume the Born-Oppenheimer
approximation only for the nuclear motion in the electronic ground
state of the $N$-electron molecule, which is physically reasonable for
many molecules. However, this neither implies  adiabaticity for the
scattering electron nor for excited states of the scattering
complex. By way of contrast, the nonadiabatic coupling of the
projectile motion to the nuclear motion, which can be very important
for slow projectiles, is explicitly accounted for in our formalism.

Defining the operator $Q = 1 - P$ as a projector onto the secondary
space ($1$ is here the identity operator in ${\sf Y}$-space
\cite{note-Y-space}), we can easily partition Eq.~(\ref{umf2}) in
order to obtain an effective equation in the primary space. We take
the following steps:

Inserting the identity $1 = P + Q$ into Eq.~(\ref{umf2}) yields
\beq \label{umf3}
  \Hb (P+Q)  \left( \kpt \atop 0 \right) = (E - \Hn + E_{00}) \left(
  \kpt \atop 0 \right) .
\eeq
Acting on this equation with $Q$ from the left and using $[Q,\,\Hn]
=0$, which follows from $[P,\,\Hn] =0$, yields an equation for the
secondary-space component of the total wavefunction $Q \left( \kpt
,\,0 \right)^t$:
\beq \label{umf4}
  \Hb_{QP}\, P \left( \kpt \atop 0 \right) = ( E - \Hn + E_{00}- \Hb_{QQ})
  \, Q \left( \kpt \atop 0 \right) .
\eeq
Here we introduced the notation $\Hb_{QP}$ for the operator product $Q
\Hb P$, etc. 
Using Eq.~(\ref{umf4}) to replace the secondary-space component from the
$P$-projection of Eq.~(\ref{umf3}) leads to the desired equation for
the primary-space component of the wavefunction:
\beq \label{umf5}
  \left[ \Hb_{PP}+ \Hn - E_{00} - \Hb_{PQ} \frac{1}{E - \Hn + E_{00}- 
  \Hb_{QQ}} \Hb_{QP} \right] P \left( \kpt \atop 0 \right) = E P \left( \kpt
  \atop 0 \right) .
\eeq
This is the resulting form of the time-independent Schr\"odinger
equation projected to the primary space. The projected component
fulfils a pseudo-eigenvalue equation with an energy-dependent operator
on the left-hand side. This operator acts on the nuclear coordinates
$\R$ and on the electronic coordinates, but only within the primary
space, which is isomorphous to the one-particle space (as was
discussed in Sec.~\ref{ssec:trad}).

In the following we will make contact to our previous formulation and
identify the self energy of the dynamical Green's function as the
optical potential.  The primary-space component of the wavefunction is
given by
\beq
   P \left( \kptR \atop 0 \right) = \sum_q \ky{q} \left[ \By{q} \left(
   \kptR \atop 0 \right) \right] = \sum_q \ky{q}  f_q(\R),
\eeq
where $f_q(\R) = \bpsi a_q \kptR$ is the effective wavefunction of
Eq.~(\ref{def-scat-ampl}) in the orbital representation. Taking the
inner product of Eq.~(\ref{umf5}) with $\by{p}$ from the left yields
\beq \label{umf6}
  \nonumber {
  \sum_q \left( \by{p} \Hb \ky{q} + (\Hn - E_{00}) \delta_{pq} - \by{p} \Hb Q
  \frac{1}{E - \Hn + E_{00}- \Hb_{QQ}} Q \Hb \ky{q} \right) f_q(\R)}\\
  = E f_p(\R) .
\eeq
The first term on the left can be identified as the primary block
$\mat{H}_{aa}$ of Eq.~(\ref{ev-Hb-aa}), which splits into the zeroth order
energy matrix $\mat{\varepsilon}$ and the static
(i.~e.~energy-independent) part of the self energy
$\mat{\Sd}(\infty)$ according to Eq.~(\ref{self-energy}). Also the
energy-dependent part of $\mat{\Sd}(\omega)$ can be identified in
Eq.~(\ref{umf6}) when the representation 
\beq
  Q = {\sum_J}' \ket{Q_J}\bra{Q_J}
\eeq
is chosen for the projector to the secondary space. The sum
$\sum'$ runs over the basis states $\ket{Q_J}$ of the secondary space
only [c.~f.~Eqs.~(\ref{def-mat-Hb}) and (\ref{struc-mat-Hb})]. We obtain
\beq \label{gen_eff_eq}
  \sum_q \left[ \varepsilon_{pq}  + (\Hn - E_{00}) \delta_{pq} +
  {\Sd}_{pq}(\omega) \right] f_q(\R) = E f_p(\R) .
\eeq
The effective Schr\"odinger equation in the form of
Eq.~(\ref{eff-schroed-eq}) is obtained when using the coordinate
representation instead of the orbital representation for the
scattering electron and realising
that the matrix $\mat{\varepsilon}$ is the orbital representation of
the operator $\Hez$ of Eq.~(\ref{def-Hez}) in the one-electron
space. As discussed earlier in this subsection, the projection
operators $P$ and $Q$ are completely independent of the choice of the
single-particle basis and thus Eq.~(\ref{gen_eff_eq}) holds in the
momentum, in the coordinate-space, and in arbitrary orbital
representations. We are also not restricted to a particular choice of
the zeroth-order electronic Hamiltonian $\Hez$, represented by the
matrix $\mat{\varepsilon}$ in Eq.~(\ref{gen_eff_eq}). Any
single-particle operator, which may or may not depend on $\R$, is
possible. A convenient choice for the electron-molecule scattering
problem is certainly to choose the Hartree-Fock operator for $\Hez$
because this simplifies the calculation of the self energy
$\Sd(\omega)$. In this case, $\Hez$ contains the electronic kinetic
energy as well as the static exchange potential.

So far, we have proven that the projection $f(\r,\R)$ of
Eq.~(\ref{def-scat-ampl}) of the physical scattering wavefunction
fulfils the effective Schr\"odinger equation
(\ref{eff-schroed-eq}). There are also unphysical solutions of this
equation due to the introduction of the second component in
Eq.~(\ref{umf2}). The unphysical component lives in the Hilbert space
of $N-1$ electrons, coupled to the same nuclear degrees of freedom as
the physical component. Since the Hamiltonian for the second
component from Eq.~(\ref{umf2}) is given by
\beq
  \Vo(\R) - \He + \Hn - E_{00} = -[ \He - \Tn - 2 \Vo(\R) + E_{00}] ,
\eeq
the nuclear dynamics is not treated in the correct way to describe an
ionised molecule because $\He$ and $\Tn$ have opposite sign. If we want
to apply the formalism of the dynamical optical potential to describe
the dynamics of ionised molecules, the definition of the Hamiltonian
has to be changed accordingly. 

Coming back to the description of the electron-scattering problem we
want to consider the question of how to identify unphysical solutions
and whether the unphysical component can interfere with the physical
solutions. The answer to these questions is easily given when the
scattering boundary conditions are obeyed correctly:
Since the electronic ground state of the scattering
target $\kpsi$ represents a bound state, any overlap
$\bra{\Psi}\psi(\r)\kpsi$ is a square-integrable function of $\r$
vanishing asymptotically for $|\r| \to \infty$. Consequently, the same
holds true for the effective wavefunction $f^{\text{unphys}}(\r,\R)$
belonging to an unphysical solution of Eq.~(\ref{umf2})
\beq
  f^{\text{unphys}}(\r,\R) = \left[ \Bra{Y(\r)} \left(0 \atop
   \bra{\Psi^{N-1}_{\text{unphys}}(\R)} \right) \right] 
  = \bra{\Psi^{N-1}_{\text{unphys}}(\R)} \psi(\r)\kpsi ,
\eeq
as a function of $\r$.
As long as one solves the effective Schr\"odinger equation
(\ref{eff-schroed-eq}) or the corresponding Lippmann-Schwinger
equation imposing scattering boundary conditions on the effective
wavefunction $f(\r,\R)$, one obtains a physical solution. Even a
contamination of the physical solution with a short-range unphysical
solution will do no harm because for the calculation of the
$S$-matrix elements only the asymptotic behaviour of the effective
wavefunction enters.

Let us briefly examine the only approximation made in
the derivation of Eq.~(\ref{umf5}). Without the approximation
(\ref{approx-comm}), which allowed the commutation of the nuclear
kinetic energy $\Tn$ with the projection operators $P$ and $Q$,
additional terms proportional to $P\Hn Q$ and $Q\Hn P$ appear in the
projected Schr\"odinger equation. These terms describe virtual
excitations from the primary to the secondary space, which mean an
electronic excitation in the target molecule, mediated by the
nuclear kinetic energy operator. These terms compete with excitations
by the electronic Hamiltonian $P \He Q$ and $Q \He P$ against which
they can usually be neglected. The only exemption may be given in
cases where the Born-Oppenheimer approximation for the electronic
ground state fails and the potential surfaces of the electronic ground
and excited states of the target molecule come close to each other. In
these rare cases, the vibronic interactions have to be accounted for
and a diabatic representation for the ground state may be
adequate. The dynamical optical potential then has to be augmented by
terms that describe the corresponding vibronic transitions.


\subsection{Approximative scattering equations}
\label{sec:approximate-Tn} 

In this subsection we want to discuss 
different approximations to the exact scattering equations and the
physical models these approximations imply. In the
first two points we will consider
possibilities to regain well-known approximate descriptions of
electron-molecule scattering by either neglecting the energy-dependent
optical potential or by freezing the nuclear degrees of freedom in the
exact equation (\ref{schr-eq-Layzer}). In the last two points we will
concentrate on the dynamical optical potential itself. Apart from
discussing the role of the nuclear dynamics within the optical
potential we will show how to compute {\it ab initio} dynamical
optical potentials. 

\subsubsection{Static approximations to the optical potential}
\label{sec:static-appr} 

First of all we want to consider the simplest of all approximations to
the optical potential, which results by neglecting the optical
potential ${\Sd}(\omega)$ altogether. Formally this approximation is
equivalent to neglecting the interaction part $\Hei$ in the electronic
Hamiltonian $\He$ of Eq.~(\ref{part-He}). This equivalence follows
easily from Eq.~(\ref{Sd-by-Sigma}) which connects the dynamical self
energy ${\Sd}(\omega)$ to the usual self energy ${\Sigma}(\omega)$,
which is by definition at least of first order in $\Hei$
\cite{fetter71}. In other words we can say  that, neglecting the
dynamical self energy ${\Sd}(\omega)$, the effective Schr\"odinger
equation (\ref{eff-schroed-eq}) is still exact for physical systems
where the scattering electron is not correlated with other
electrons in the system and thus is well described by $\Hez$. The
one-particle operator $\Hez$ may still contain mean fields or forces
exerted by the atomic nuclei or external fields, of course.

The level of approximation gained with the resulting static
(i.~e.~energy-independent) Layzer operator ${\cal L}^{\text{st}} = \Hez +
\Hn - E_{00}$ depends on the choice of the zeroth-order electronic
Hamiltonian $\Hez$:
\begin{itemize}
\item Choosing $\Hez$ to describe the electronic kinetic energy yields
  a Layzer operator ${\cal L}^{\text{st}}$ that describes the separable
  motion of a free electron and a vibrating or rotating target
  molecule in its electronic ground state.

\item Choosing instead $\Hez$ to describe the $\R$-dependent
  Hartree-Fock operator yields a static Layzer operator that describes the
  motion of the scattering electron under the influence of the static
  and the exchange interactions with the Hartree-Fock charge cloud of
  the target molecule coupled to the nuclear motion in the usual
  electronic-ground-state potential $\Vo(\R)$ augmented by the
  Coulomb attraction between the nuclei and the electronic projectile. This
  approximation gives a consistent treatment of the electron-molecule
  scattering problem in the strict single-particle picture. Projectile
  and nuclear motion are fully coupled but the very nature of this
  static approximation excludes polarisation effects as well as the
  possibility of electronic excitation of the target caused by the
  impact of the scattering projectile. Another defect of this
  Hartree-Fock based static-exchange approximation is that the static
  charge cloud of the target molecule is only described in an
  approximate (uncorrelated) manner through the Hartree-Fock
  wavefunction.
\end{itemize}

Another straightforward possibility for an energy-independent
approximation to the dynamical optical potential is to include the
high-energy limit of the self energy $\Sd(\infty)$, which is also
called the static self energy. The static part of the dynamical self
energy $\Sd(\infty)$ is identical to the purely-electronic static self
energy $\Sigma(\infty)$, as can be seen from
Eqs.~(\ref{elec-self-energy}) and (\ref{self-energy}) [see also
Eq.~(\ref{Sd-by-Sigma})]. As shown in Ref.~\cite{weikert87}, this nonlocal
operator can be interpreted to improve the static-exchange interaction
beyond the Hartree-Fock description of the target wavefunction
including target correlation. The present approximation to the Layzer
operator ${\cal L}^{\text{cse}} = \Hez +
\Hn + \Sd(\infty) - E_{00}$ is independent of the choice of the
zeroth-order electronic Hamiltonian $\Hez$ in Eq.~(\ref{part-He})
because the static self energy takes account of the interaction term
$\Hei$. The Layzer operator $\cal{L}^{\text{cse}}$ corresponds to a
scattering potential that originates from a correlated but static
charge distribution of the target molecule, which is also known as the
correlated static exchange (cse) potential.

Approximations to the energy-dependent part of the dynamical optical
potential in the context of the fully coupled scattering problem will
be discussed in Secs.~\ref{sec:expansion-Tn} and
\ref{sec:exp-opt-pot-Hn}. In the following paragraph we will consider
the simplified case where the nuclear dynamics is neglected
altogether.

\subsubsection{Fixed-nuclei scattering} \label{sec:fixed-nuclei}

Let us recall that the full Layzer operator $\cal{L}$ may be expressed
with the help of Eq.~(\ref{Sd-by-Sigma}) by
\beq
  {\cal{L}}(E) = \Hez + \Tn + \Vo(\R) + {\Sigma}(E -\Tn - \Vo(\R) +
  E_{00})  - E_{00} . 
\eeq
In the limit of infinitely heavy nuclei, the nuclear kinetic energy
$\Tn$ can be neglected. We call this limit the fixed-nuclei (fn) limit
because the Layzer operator
\beq
  {\cal{L}^{\text{fn}}}(E) = \Hez + {\Sigma}(E -
  \Vo(\R) +  E_{00})  + \Vo(\R) - E_{00} 
\eeq
and the corresponding scattering wavefunction $f(\r,\R)$ of
Eq.~(\ref{eff-schroed-eq}) now depend only parametrically on the
nuclear coordinates $\R$. The operator ${\cal{L}^{\text{fn}}}(E)$ is
exactly the Layzer operator used in the literature for electron
scattering from atoms or rigid molecules
\cite{bell59,layzer63,csanak71} apart from the $\R$-dependent energy shift
$\Vo(\R) - E_{00}$. This energy shift resets the zero point of the
energy scale from $E_{00}$ to $\Vo(\R)$.

In the present case of fixed-nuclei scattering, the optical potential
is given by the purely electronic self energy $\Sigma(\omega)$. As
mentioned earlier, the static part $\Sigma(\infty)$ improves the
static-exchange potential with respect to the correlation of the
targets electronic ground state wavefunction. The energy-dependent
part $M(\omega) = \Sigma(\omega) - \Sigma(\infty)$ is, according to
Eq.~(\ref{elec-self-energy}), given by
\beq \label{elec-dyn-self-energy}
  \mat{M}(\omega) = 
  \mat{\Hb}_{ab} \frac{1}{\omega \mat{1} - \mat{\Hb}_{bb}}
  \mat{\Hb}_{ba} .
\eeq
For convenience, we again resort to the matrix notation for electronic
coordinates used already in the preceding chapters. Of course, the
matrices can also be expressed in the coordinate
representation. ${M}(\omega)$ accounts for the so-called dynamic
correlation including the polarisation of the target by the incident
projectile electron \cite{meyer88}.  The energy dependence of
$M(\omega)$ has been studied in the literature
\cite{cederbaum77} and the consequences of the energy dependence of
optical potentials for scattering systems have been discussed
\cite{feshbach58}. Here, we briefly want to remark the following:

$M(\omega)$ introduces poles and branch cuts in the optical potential,
which can be found by diagonalising the denominator of
Eq.~(\ref{elec-dyn-self-energy}). The matrix $\mat{\Hb}_{bb}$ in the
denominator is already diagonal in zeroth order, e.~g.~in the
Hartree-Fock approximation. The diagonal elements correspond to
electronic excitations of the $(N+1)$- and $(N-1)$-electron systems
and can be classified by configurations of $2p-h$, $3p-2h$,
etc.~character as well as configurations of $p-2h$, $2p-3h$ type, and
so on. Explicit matrix representations for $\mat{\Hb}_{ab}$,
$\mat{\Hb}_{ba}$, and $\mat{\Hb}_{bb}$ can be found in
Ref.~\cite{aTarantelli92}.

For the calculation of $\Sigma(\omega)$, standard
approximation schemes are available. The so-called algebraic
diagrammatic construction (ADC) scheme \cite{schirmer83}, for example,
yields a perturbation-theory-based hierarchy of approximations for
$\Sigma(\omega)$ that preserve the analytic structure of this
function of $\omega$, in contrast to ordinary perturbation theory,
which does not. The ADC approximation to $\Sigma(\omega)$ has the
structure of  Eq.~(\ref{elec-self-energy}) but the matrices
$\mat{\Hb}_{ij}$ are constructed in the $n$th order ADC scheme such
that the perturbation expansion of the approximate
$\Sigma^{\text{ADC}}(\omega)$ with respect to the electronic
interaction $\Hei$ of Eq.~(\ref{part-He}) coincides up to $n$th order
with the usual Feynman-Dyson perturbation series of $\Sigma(\omega)$.

Since the poles and cuts relating to the ionised molecule
($N-1$ electrons) appear at negative energies, they only have a weak
influence on the energy-dependence of the optical potential at the
relevant scattering energies.
The poles and cuts at positive energies, however, can be associated
with excited states of the $(N+1)$-electron scattering
system. Signatures of Feshbach resonances therefore can be found in
the dynamic part of the optical potential although the exact resonance
position and width, of course, have to be calculated from the full
effective Hamiltonian. A branch cut will appear in the analytic
structure of $\Sigma(\omega)$ above the first excitation energy of the
target molecule because with sufficient energy for an electronic
excitation, a new channel opens and inelastic scattering becomes
possible. The branch cut of $\Sigma(\omega)$ has the consequence that
the optical potential acquires an imaginary component and becomes
non-hermitian, thereby accounting for the loss of amplitude in the
channel of electronically elastic scattering. 

\subsubsection{Nuclear dynamics in the optical potential: Expansion with
respect to $\Tn$} \label{sec:expansion-Tn}

We are now going to discuss the influences of the
nuclear motion on the dynamical optical potential $\Sd(E)$.
The dynamical self energy $\Sd(E)$ as given by Eq.~(\ref{self-energy})
consists of an energy-independent (static) part $\Sd(\infty)$ and an
energy-dependent part $\Md(E) = \Sd(E) - \Sd(\infty)$, which vanishes
for large energies $E$.
The energy-dependent part
$\Md(E)$ includes all effects of dynamic correlation including
polarisation but is now modified with respect to the fixed-nuclei case
considered in Sec.~\ref{sec:fixed-nuclei}.
As follows from Eq.~(\ref{self-energy}), the energy-dependent part
reads in matrix notation:
\beq \label{en-dep-dyn-self-energy}
  \mat{\Md}(E) = \mat{\Hb}_{ab} \frac{1}{[E - \Tn
  -\Vo(\R) + E_{00}] \mat{1} - \mat{\Hb}_{bb}}  \mat{\Hb}_{ba} .
\eeq
Like before, $\Tn$ denotes the nuclear kinetic energy, $\Vo(\R)$ the
(electronic) ground-state potential, $E$ is the scattering energy, and
$E_{00}$ the molecular ground-state energy. The matrices
$\mat{\Hb}_{ij}$ are the partial matrix representations of the
two-component electronic excitation-energy operator $\Hb$, like
introduced in Sec.~\ref{ssec:trad}, and depend on the nuclear
coordinates $\R$.  In order to diagonalise the denominator to find the
pole structure of ${\Md}(E)$, not only eigenvectors of
$\mat{\Hb}_{bb}$ corresponding to electronic eigenstates in the
secondary space are necessary but instead one also has to consider the
additional degrees of freedom of nuclear motion. Since the energies
usually associated to the nuclear dynamics are much smaller than
typical energies of electronic excitations, we may assume that the
coarse structure of the energy dependence does not differ much from
the fixed-nuclei case discussed in
Sec.~\ref{sec:fixed-nuclei}. However, the nuclear dynamics will
introduce a fine structure and may become very important, either at
energies where electronic excitation is possible or nearly possible,
or in situation that are very sensitive to disturbance of the
scattering potential, e.~g.~close to vibrationally inelastic
thresholds.

As long as the scattering energies are safely
away from Feshbach resonances or inelastic thresholds, the influences
of nuclear dynamics on the optical potential itself may either be neglected
or accounted for approximately by expansions of the denominator in
Eq.~(\ref{en-dep-dyn-self-energy}). We will now proceed to discuss two
convenient expansions for $\mat{\Md} (E)$.

Closest in spirit to the before-discussed fixed-nuclei approximation
is the assumption that $\Tn$ represents a small perturbation,
suggesting the following type of expansion:
\beq \label{Md-expansion-Tn}
  \mat{\Md} (E) = \mat{\Hb}_{ab} \frac{1}{[E 
  -\Vo(\R) + E_{00}] \mat{1} - \mat{\Hb}_{bb}} \sum_{\nu = 0}^\infty
  \left( \Tn\frac{1}{[E 
  -\Vo(\R) + E_{00}] \mat{1} - \mat{\Hb}_{bb}} \right)^\nu \mat{\Hb}_{ba} .
\eeq
In the present context, we call the first term of this expansion ($\nu
= 0$) the {\bf adiabatic optical potential (aop)}:
\beqs \label{adiab-Sd}
  \nonumber
  \mat{\Sd}^{\aop} (E) & = & \mat{\Hb}_{aa} -
  \mat{\varepsilon} + \mat{\Hb}_{ab} \frac{1}{[E 
  -\Vo(\R) + E_{00}] \mat{1} - \mat{\Hb}_{bb}} \mat{\Hb}_{ba} \\
  & = & \mat{\Sigma}(E - \Vo(\R) + E_{00}).
\eeqs

Formally, the adiabatic optical potential ${\Sd}^{\aop}(E)$
coincides with the optical potential used in fixed-nuclei scattering
in Sec.~\ref{sec:fixed-nuclei}. The difference is that the nuclear
coordinates $\R$ are now dynamical variables. The adiabatic optical
potential ${\Sd}^{\aop}(E)$ is a local operator on the nuclear
coordinates $\R$ because $\Vo$ and the matrices $\mat{\Hb}_{ij}$
depend on $\R$ but there are no derivatives with respect to $\R$ in
${\Sd}^{\aop}(E)$. Since ${\Sd}^{\aop}(E)$ can be
calculated, for each $\R$, from the purely electronic self energy
$\Sigma(E)$, standard techniques for calculating the electronic self
energy like the ADC method mentioned in Sec.~\ref{sec:fixed-nuclei}
can be applied.

The  Layzer operator corresponding to the presently discussed
approximative treatment of the dynamical optical potential ${\Sd}^{\aop}(E)$
is given by
\beq
  {\cal L}^{\aop}(E) = \Hez + \Hn + {\Sd}^{\aop}(E)
  - E_{00} .
\eeq
Although the dynamical optical potential is treated here in an
adiabatic approximation, the Layzer operator ${\cal
L}^{\aop}(E)$ describes the fully and non-adiabatically coupled motion of the
projectile electron and the atomic nuclei with a simplified optical
potential. In this approximation, the nuclear dynamics in the
electronically excited states of the target is treated adiabatically.
Even though this approximation may not be justified when describing
experiments probing in detail the electronic excitation structure of
the target, the situation is different for scattering at low but not
too low energies. When electronic excitations are forbidden by energy
conservation because the electron-impact energy is too low, the
adiabatic approximation ${\Sd}^{\aop}(E)$ may be very good. At low
impact energies, electronic excitation then is still possible as a
virtual process leading to polarisation of the molecule but obeying
the energy-time uncertainty relation. When the typical times for these
virtual excitations are much smaller than the time scales of nuclear
vibrations, which will usually be the case, then the fixed-nuclei
approximation for the dynamical optical potential ${\Sd}^{\aop}(E)$ is
adequate.

For very slow scattering electrons, on the other hand, it may also be
necessary to improve the approximation ${\Sd}^{\aop}(E)$, when
the cross sections probe the details of the optical potential very
sensitively. In the adiabatic
optical potential ${\Sd}^{\aop}(E)$, the dynamical relaxation
of the nuclear structure during the polarisation of the electronic
charge cloud induced by the incoming projectile is not treated
properly. At very low projectile velocities where the trajectories are
most sensitive on the coupling to the nuclear motion this may become
an important defect of the theory. In this case the expansion of the
dynamical optical potential discussed in the next subsection, which
allows for an approximate treatment of the neglected effects, becomes
most valuable.

\subsubsection{Expansion of the dynamical optical potential with
respect to $\Hn-E_{00}$} \label{sec:exp-opt-pot-Hn}

The nuclear kinetic energy operator $\Tn$ introduces derivatives with
respect to the nuclear coordinates $\R$ in the terms of the expansion
(\ref{Md-expansion-Tn}) for $\nu \ge 1$.  In general, these terms will
be difficult to evaluate because the approximate and also the exact
matrices $\mat{\Hb}_{ij}$ may depend strongly on the nuclear
coordinates $\R$ due to the electron-nucleon Coulomb repulsion
contained in $\Hb$. Another expansion that is better suited for
higher-order approximations is obtained when taking $\Tn + \Vo(\R) -
E_{00} = \Hn - E_{00}$ as the ``small'' perturbation. The assumption
that $\Hn - E_{00}$ may be regarded small in the denominator of
Eq.~(\ref{en-dep-dyn-self-energy}) is justified when the energies
associated to nuclear excitations appearing during the scattering
process and possibly during (virtual) electronic excitations are small
compared to the electronic excitation energies of the target. The
corresponding expansion of the energy-dependent part of the dynamical
self energy reads
\beq \label{Md-expansion-Hn}
  \mat{\Md}(E) =\mat{\Hb}_{ab} \frac{1}{E 
   \mat{1} - \mat{\Hb}_{bb}} \sum_{\nu = 0}^\infty
  \left[ (\Hn - E_{00}) \frac{1}{E
   \mat{1} - \mat{\Hb}_{bb}} \right]^\nu \mat{\Hb}_{ba} .
\eeq
The first term $\mat{\Md}^0(E)$ of this expansion ($\nu = 0$) can be
identified as the energy-dependent (dynamic) part of the purely
electronic self energy (\ref{elec-dyn-self-energy}) and will be called
the {\bf zero-point optical potential} because it reflects the optical
potential of ground-state nuclear motion:
\beqs
  \nonumber
  \mat{\Md}^0(E) & = & \mat{\Hb}_{ab} \frac{1}{E 
   \mat{1} - \mat{\Hb}_{bb}}\mat{\Hb}_{ba}
   = \mat{M}(E) 
\eeqs
The zero-point optical potential is a local operator with respect to
the nuclear coordinates $\R$ like the adiabatic optical potential
$\mat{\Md}^{\aop}(E)$, which was discussed above. 

The higher order terms in the expansion for $\nu \ge 1$, however, are
differential operators in the nuclear coordinates. They can be
transformed into nonlocal integral operators $\mat{\Md}^{\nu}(E)\,
\underline{f}(\R) = \int d\R'\,\mat{\Md}^{\nu}(E,\R,\R')\,
\underline{f}(\R')$ by inserting the resolution of the identity with
respect to nuclear motion $\sum_k
\ket{\phi_k}\bra{\phi_k}$. In the coordinate representation for the
nuclear degrees of freedom $\R$, the integral kernel belonging to the
second term of the expansion (\ref{Md-expansion-Hn}) reads
\beq \label{Md1}
  \mat{\Md}^1(E,\R,\R') = \sum_k \mat{\Hb}_{ab}(\R) \frac{1}{E 
   \mat{1} - \mat{\Hb}_{bb}(\R)}\phi_k(\R)[E_{0k} -
   E_{00}]\phi^*_k(\R') \frac{1}{E \mat{1} - \mat{\Hb}_{bb}(\R')}
   \mat{\Hb}_{ba}(\R') ,
\eeq
where we explicitely indicated the nuclear-coordinate dependence of
the matrices $\mat{\Hb}_{ij}$. The numerical realisation of this
approximation is not very difficult because the nuclear eigenvalues
$E_{0k}$ and functions $\phi_k(\R)$ can be calculated easily form the
ground-state potential $\Vo(\R)$ and approximate matrices
$\mat{\Hb}_{ij}$ are available, e.~g., by the ADC approximation as
described above in Sec.~\ref{sec:fixed-nuclei}.

Eq.~(\ref{Md1}) is remarkable because it allows us to estimate the
importance of the higher-order terms in the expansion
(\ref{Md-expansion-Hn}). If the nuclear dynamics of the target
molecule, in spite of virtual excitations by the scattering electron,
can be approximated well by the zero-point vibration of the isolated
molecule (in its electronic ground state) then all terms with $k \ne
0$ in Eq.~(\ref{Md1}) can be neglected. Since, however, the energy
difference $[E_{0k} - E_{00}]$ vanishes for $k=0$, the term
$\mat{\Md}^1(E)$ and the higher order terms in (\ref{Md-expansion-Hn})
vanish. In this approximation only the zeroth term $\mat{\Md}^0(E)$
contributes in the expansion of the dynamical optical potential. In
other words $\mat{\Md}(E)$ is a local operator in the nuclear
coordinates if the nuclear-coordinate dependence of the effective
scattering wavefunction $f(\r,\R)$ is given by the nuclear
eigenfunction $\phi_0(\R)$ and the $\R$ dependence of the matrices
$\mat{\Hb}_{bb}$ and $\mat{\Hb}_{ba}$ can be neglected. This will not
always be the case. However, it can also be tested during a
calculation for which values of ${k}$ the integral
\beq
  \int d\R\, \phi^*_{{k}}(\R) \frac{1}{E \mat{1} -
   \mat{\Hb}_{bb}(\R)} \mat{\Hb}_{ba}(\R) \, \underline{f}(\R)
\eeq
gives the largest contribution. The approximation $\mat{\Md}^0(E)$ can
then be improved by expanding around $\Hn- {\tilde{E}}$, where
${\tilde{E}}$ is a suitably chosen mean vibrational (rotational) energy,
instead of $\Hn-E_{00}$, yielding the zeroth approximation
\beq
  \mat{\Md}^{\tilde{0}}  = \mat{\Sigma}(E+ E_{00} - {\tilde{E}} ) -
  \mat{\Sigma}(\infty) 
\eeq
and the corresponding higher approximations like in
Eq.~(\ref{Md-expansion-Hn}). 

The difference between the adiabatic optical potential
${\Sd}^{\aop}(E)$ and the zero-point optical potential
$\Sd^0(E)$ or the improved approximation $\Sd^{\tilde{0}}(E)$ lies in
the energy reference of the denominator of the optical potential. We
expect $\Sd^0(E)$ or $\Sd^{\tilde{0}}(E)$ to be suitable
approximations for $\Sd(E)$ when the nuclear motion is more or less
confined to the immediate surroundings of the equilibrium
configuration $\R_{\text{eq}}$ of the molecule. In particular for
processes that travel between different levels on the potential energy
surface $\Vo(\R)$ corresponding to a considerable transfer of energy
between the projectile and target like it happens in associative
detachment or dissociative attachment, we expect the adiabatic
approximation ${\Sd}^{\aop}(E)$ to be the better-suited
approximation. However, we want to stress once again that the treatment of
the nuclear dynamics within the optical potential is certainly less
essential than the direct coupling between projectile and nuclear
motion in the effective Schr\"odinger equation (\ref{eff-schroed-eq}),
as long as the scattering energy remains well below the threshold for
electronic excitation of the molecular target.
The expansion (\ref{Md-expansion-Hn}) presents a systematic
possibility to improve upon the adiabatic or the zero-point
approximation that is applicable if the scattering process
has too little energy to access the electronically excited states of
the target. The first-order term $\mat{\Md}^{1}$ of Eq.~(\ref{Md1})
and the corresponding higher approximations provide access to a
systematic improvement of the discussed approximations for {\it ab
initio} methods of quantum chemistry that may prove necessary to
describe the dynamical couplings correctly for very-low-energy scattering
processes that are particularly sensitive to the correct description
of the optical potential.
If the scattering energy is above or close to the threshold for
electronic excitation of the target molecule, the effective Schr\"odinger
equation (\ref{eff-schroed-eq}) still remains valid but other
approximations to the dynamical optical potential have to be used.

\section{Conclusions}

In this paper we have derived a rigorous optical potential for the
coupled motion of the projectile electron and the atomic nuclei in
electron-molecule scattering. The dynamical optical potential should
be particularly valuable for studying the non-adiabatic coupling of
projectile and nuclear motion close to inelastic thresholds where the
projectile velocity is comparable to the typical velocities of nuclear
motion. Our work extends the well-known many-body formalism of optical
potentials beyond the fixed-nuclei approximation. On the other hand
the rigorous derivation and the explicit expressions given for the
dynamical optical potential provide a justification and prospective for
enhancement of model polarisation potentials used commonly in
close-coupling and $R$-matrix calculations.

The dynamical optical potential can describe all kinds of
electronically elastic scattering processes such as elastic
scattering, vibrational or rotational excitation (de-excitation),
dissociative attachment, and associative detachment. Also when
electronic excitations are energetically accessible, the dynamical
optical potential may be used to calculate the (electronically) elastic
partial cross sections as well as resonance positions and widths.

One central result of the present work is that for scattering energies
far enough away from electronic excitations of the target, the
dynamical optical potential is given in a first approximation by the
static-exchange potential augmented by the fixed-nuclei self energy
$\Sigma(\omega)$ of the purely electronic Green's function. This
approximation, which yields a local operator with respect to the
nuclear coordinates $\R$, can be improved by a hierarchy of non-local
terms. Due to the close relation to the usual self energy
$\Sigma(\omega)$, all terms can be calculated with standard {\em
ab-initio} methods using well-defined approximation schemes. 

The dynamical optical potential provides a firm basis for theoretical
studies of coupled projectile and nuclear motion in ultra-low-energy
electron-molecule scattering, which is becoming a vivid field of
interest due to recent experimental advances (see
e.~g.~\cite{schramm98-liste}).  The effective scattering equations presented
in this paper can be solved, e.~g.~by using close-coupling expansions
or other time-independent techniques. We expect that particularly
valuable insight into the mechanisms of the scattering process,
however, can be gained by time-dependent wave-packet calculations.
Solving these scattering equations certainly remains a numerically
demanding task but becomes more and more feasible now with state of
the art computing facilities and the advancement of numerical methods
\cite{cerjan93}.

\section*{Acknowledgements}

We would like to thank E.~Pahl for a critical reading of the
manuscript and helpful comments. Financial support by the Deutsche
Forschungsgemeinschaft (DFG) through the Forschergruppe
``Schwellenverhalten, Resonanzen und nichtlokale Wechselwirkungen bei
niederenergetischen Elektronenstreuprozessen'' is gratefully
acknowledged.




\begin{thebibliography}{10}

\bibitem{feshbach92}
H. Feshbach, {\em Theoretical Nuclear Physics: Nuclear Reactions} (Wiley, New
  York, 1992).

\bibitem{feshbach58}
H. Feshbach, Ann. of Phys. (N.Y.) {\bf 5},  357  (1958).

\bibitem{feshbach62}
H. Feshbach, Ann. of Phys. (N.Y.) {\bf 19},  287  (1962).

\bibitem{bell59}
J.~S. Bell and E.~J. Squires, Phys. Rev. Lett. {\bf 3},  96  (1959).

\bibitem{csanak71}
G.~Y. Csanak, H.~S. Taylor, and R. Yaris, Adv. Mol. Phys. {\bf 7},  287
  (1971).

\bibitem{ring80}
P. Ring and P. Schuck, {\em The Nuclear Many--Body Problem} (Springer, New
  York, 1980), p.\ 623 ff.

\bibitem{Mahaux91}
C. Mahaux and R. Sartor, Nucl. Phys. A {\bf 530},  303  (1991).

\bibitem{schneider82}
B.~I. Schneider and L.~A. Collins, J. Phys. B {\bf 15},  L 335  (1982).

\bibitem{schneider83}
B.~I. Schneider and L.~A. Collins, Phys. Rev. A {\bf 27},  2847  (1983).

\bibitem{lengsfield91}
B.~H. {Lengsfield III} and T.~N. Rescigno, Phys. Rev. A {\bf 44},  2913
  (1991).

\bibitem{gil93}
T.~J. Gil, C.~W. McCurdy, T.~N. Rescigno, and B.~H. {Lengsfield III}, Phys.
  Rev. A {\bf 47},  255  (1993).

\bibitem{fetter71}
A.~L. {Fetter} and J.~D. Walecka, {\em Quantum Theory of Many--Particle
  Systems}, 1st ed. (McGraw--Hill, New York, 1971).

\bibitem{schneider70}
B. Schneider, H.~S. Taylor, and R. Yaris, Phys. Rev. A {\bf 1},  855  (1970).

\bibitem{schirmer83}
J. Schirmer, L.~S. Cederbaum, and O. Walter, Phys. Rev. A {\bf 28},  1237
  (1983).

\bibitem{ortiz88}
J.~V. Ortiz, J. Chem. Phys. {\bf 89},  6348  (1988).

\bibitem{klonover77}
A. Klonover and U. Kaldor, Chem. Phys. Lett. {\bf 51},  321  (1977).

\bibitem{klonover78}
A. Klonover and U. Kaldor, J. Phys. B {\bf 11},  1623  (1978).

\bibitem{klonover79}
A. Klonover and U. Kaldor, J. Phys. B {\bf 12},  323  (1979).

\bibitem{berman83}
M. Berman, O. Walter, and L.~S. Cederbaum, Phys. Rev. Lett. {\bf 50},  1979
  (1983).

\bibitem{meyer89}
H.-D. Meyer, Phys. Rev. A {\bf 40},  5605  (1989).

\bibitem{wigner47}
E.~P. Wigner and L. Eisenbud, Phys. Rev. {\bf 26},  29  (1947).

\bibitem{gillan87-liste}
{C.~J.~Gillan, O.~Nagy, P.~G.~Burke, L.~A.~Morgan, and C.~J.~Noble}, J. Phys. B
  {\bf 20},  4585  (1987).

\bibitem{pfingst94}
K. Pfingst, B.~M. Nestmann, and S.~D. Peyerimhoff, J. Phys. B {\bf 27},  2283
  (1994).

\bibitem{lucchese86}
R.~R. Lucchese, K. Takatsuka, and V. McKoy, Phys. Reports {\bf 131},  147
  (1986).

\bibitem{miller87}
W.~H. Miller and B.~M. D.~B. {Jansen op de Haar}, J. Chem. Phys. {\bf 86},
  6213  (1987).

\bibitem{mccurdy87}
C.~W. McCurdy, T.~N. Rescigno, and B.~I. Schneider, Phys. Rev. A {\bf 36},
  2061  (1987).

\bibitem{lane80}
N.~F. Lane, Rev. Mod. Phys. {\bf 52},  29  (1980).

\bibitem{morrison95}
M.~A. Morrison and W. Sun,  in {\em Computational Methods for Electron-Molecule
  Collisions}, edited by W.~M. Huo and F.~A. Gianturco (Plenum Press, New York,
  1995), Chap.~6, p.\ 131.

\bibitem{case56}
D.~M. Case, Phys. Rev. {\bf 104},  838  (1956).

\bibitem{temkin67}
A. Temkin and K.~V. Vasavada, Phys. Rev. {\bf 160},  109  (1967).

\bibitem{morrison84}
M.~A. Morrison, A.~N. Feldt, and B.~C. Saha, Phys. Rev. A {\bf 30},  2811
  (1984).

\bibitem{o'connell83}
J.~K. O'Connell and N.~F. Lane, Phys. Rev. A {\bf 27},  1893  (1983).

\bibitem{morrison87}
M.~A. Morrison, B.~C. Saha, and T.~L. Gibson, Phys. Rev. A {\bf 36},  3682
  (1987).

\bibitem{morrison93}
M.~A. Morrison and W.~K. Trail, Phys. Rev. A {\bf 48},  2874  (1993).

\bibitem{morrison97}
M.~A. Morrison, W. Sun, W.~A. Isaacs, and W.~K. Trail, Phys. Rev. A {\bf 55},
  2786  (1997).

\bibitem{cederbaum81}
L.~S. Cederbaum and W. Domcke, J. Phys. B {\bf 14},  4665  (1981).

\bibitem{domcke91}
W. Domcke, Phys. Reports {\bf 208},  97  (1991).

\bibitem{cizek98}
M. \v{C}{\'{\i}}\v{z}ek, J. Hor\'a\v{c}ek, and W. Domcke, J. Phys. B {\bf 31},
  2571  (1998).

\bibitem{crawford71}
O.~H. Crawford, Mol. Phys. {\bf 40},  585  (1971).

\bibitem{note-dipole}
So far authors have only treated the $J=0$ case, ignoring the problems of
  higher angular momenta. In principle, they can be treated in the PO formalism
  but parameters for the angular momentum couplings are difficult to calculate
  directly and difficult to model as well. See, e.~g.~W.~Domcke and
  C.~M{\"u}ndel, J.~Phys.~B {\bf 18}, 4491 (1985), J.~Hor\'a\v{c}ek,
  M.~\v{C}{\'{\i}}\v{z}ek, and W.~Domcke, Theor.~Chem.~Acc., in press.

\bibitem{cederbaum96}
L.~S. Cederbaum, Few-Body Systems {\bf 21},  211  (1996).

\bibitem{aTarantelli92}
A. Tarantelli and L.~S. Cederbaum, Phys. Rev. A {\bf 45},  2790  (1992).

\bibitem{capuzzi96}
F. Capuzzi and C. Mahaux, Ann. of Phys. (N.Y.) {\bf 245},  147  (1996).

\bibitem{brand96}
J. Brand and L.~S. Cederbaum, Ann. of Phys. (N.Y.) {\bf 252},  276  (1996).

\bibitem{note-Y-space}
The vectors $\ky{p}$ may be understood as elements of a composite Hilbert space
  ${\sf Y}$ that is a direct sum of the $N+1$ particle space and the dual space
  of the $N-1$ particle space ${\sf Y} = {\sf H}^{(N+1)} \oplus {\sf
  H}^{(N-1)\ast}$ where $N$ is the number of electrons of the target molecule.
  The primary space, which is spanned by the composite states $\ky{p}$, is only
  a (small) subspace of ${\sf Y}$. The secondary space is defined by the
  orthogonal complement of the primary space in ${\sf Y}$.

\bibitem{mertins96:II}
F. Mertins, J. Schirmer, and A. Tarantelli, Phys. Rev. A {\bf 53},  2153
  (1996).

\bibitem{note-projection}
The present derivation of Dyson's equation can also be formulated without
  reference to a particular basis of the primary or secondary space with the
  help of projection operators analogous to the derivation of the optical
  potential in Sec.~\ref{sec:opt-pot-proj} (see Ref.~\cite{capuzzi96}). The
  $\R$ dependence of the projection operators is well-defined and derives only
  from the electronic ground state $\kpsi$.

\bibitem{kutzelnigg89}
W. Kutzelnigg and D. Mukherjee, J. Chem. Phys. {\bf 90},  5578  (1989).

\bibitem{layzer63}
A.~J. Layzer, Phys. Rev. {\bf 129},  897  (1963).

\bibitem{aguilar71}
J. Aguilar and J.~M. Combes, Commun. Math. Phys. {\bf 22},  269  (1971).

\bibitem{reinhardt82}
W.~P. Reinhardt, Annu. Rev. Phys. Chem. {\bf 33},  223  (1982).

\bibitem{jolicard85}
G. Jolicard and E.~J. Austin, Chem. Phys. Lett. {\bf 121},  106  (1985).

\bibitem{riss93}
U.~V. Riss and H.-D. Meyer, J. Phys. B {\bf 26},  4503  (1993).

\bibitem{sommerfeld98-liste}
{T.~Sommerfeld, U.~V.~Riss, H.-D.~Meyer, L.~S.~Cederbaum, B.~Engels, and
  H.~U.~Suter}, J. Phys. B {\bf 31},  4107  (1998).

\bibitem{mandelshtam94}
V.~A. Mandelshtam, T.~R. Ravuri, and H.~S. Taylor, J. Chem. Phys. {\bf 101},
  8792  (1994).

\bibitem{weikert87}
H.-G. Weikert and L.~S. Cederbaum, Few-Body Systems {\bf 2},  33  (1987).

\bibitem{meyer88}
H.-D. Meyer, J. Phys. B {\bf 21},  3777  (1988).

\bibitem{cederbaum77}
L.~S. Cederbaum and W. Domcke, Adv. Chem. Phys. {\bf 36},  205  (1977).

\bibitem{schramm98-liste}
{A.~Schramm, J.~M.~Weber, J.~Kreil, D.~Klar, M.-W.~Ruf, and H.~Hotop}, Phys.
  Rev. Lett. {\bf 81},  778  (1998).

\bibitem{cerjan93}
{\em Numerical Grid methods and their Application to Schr{\"o}dinger's
  Equation}, edited by C. Cerjan (Kluver Academic Publishers, Dodrecht, 1993).

\end{thebibliography}

\end{document}